\documentclass[journal=nalefd,manuscript=article,layout=twocolumn]{achemso}

\usepackage{chemformula}    
\usepackage[utf8]{inputenc} 
\usepackage[T1]{fontenc}    
\usepackage{hyperref}       
\usepackage{subfig}
\usepackage{graphicx}
\usepackage[ruled]{algorithm2e}
\usepackage{latexsym,bm,amsmath,amssymb}
\usepackage[labelsep=period,labelfont=bf]{caption}

\author{Xiao Han}
\email{hxiao@zju.edu.cn}
\affiliation[Umich]
{Department of Electrical Engineering and Computer Science, The University of Michigan, Ann Arbor, MI 48109, USA}
\alsoaffiliation[ZJU]
{Department of Information Science and Electronic Engineering, Zhejiang University, Hangzhou 310027, China}
\alsoaffiliation[Contribute]{Authors contributed equally to this work.}
\author{Ziyang Fan}
\affiliation[Umich]
{Department of Electrical Engineering and Computer Science, The University of Michigan, Ann Arbor, MI 48109, USA}
\alsoaffiliation[ZJU]
{Department of Information Science and Electronic Engineering, Zhejiang University, Hangzhou 310027, China}
\alsoaffiliation[Contribute]{Authors contributed equally to this work.}
\author{Chao Li}
\affiliation[Umich]
{Department of Electrical Engineering and Computer Science, The University of Michigan, Ann Arbor, MI 48109, USA}
\alsoaffiliation[ZJU]
{Department of Information Science and Electronic Engineering, Zhejiang University, Hangzhou 310027, China}
\author{Zeyang Liu}
\affiliation[Umich]
{Department of Electrical Engineering and Computer Science, The University of Michigan, Ann Arbor, MI 48109, USA}
\author{L.Jay Guo}
\email{guo@umich.edu}
\affiliation[Umich]
{Department of Electrical Engineering and Computer Science, The University of Michigan, Ann Arbor, MI 48109, USA}

\title{High-Freedom Inverse Design with Deep Neural Network for Metasurface Filter in the Visible}

\begin{document}

\begin{abstract}
In order to obtain a metasurface structure capable of filtering the light of a specific wavelength in the visible band, traditional method usually traverses the space consisting of possible designs, searching for a potentially satisfying device by performing iterative calculations to solve Maxwell’s equations. In this paper, we propose a neural network that can complete an inverse design process to solve the problem. Compared with the traditional method, our method is much faster while competent of generating better devices with the desired spectrum. One of the most significant advantages is that it can handle a real spectrum as well as an artificial one. Besides, our method encompasses a high degree of freedom to generate devices, ensuring their generated spectra resemble desired ones and meeting the accuracy requirements without losing practicability in the manufacturing process.

\end{abstract}

\section{Introduction}
Metasurfaces, which constructed of 2-D artificial patterns of various materials in the subwavelength scale, have received enormous attention due to its ability of unprecedented control over the intrinsic properties of light, including the amplitude\cite{cheng2015structural,proust2016all}, phase\cite{khorasaninejad2016metalenses,khorasaninejad2017metalenses}, polarization\cite{yu2012broadband,guo2016broadband} and the orbital angular momentum\cite{devlin2017arbitrary}. The most critical feature of metasurfaces is that the spatially varying patterns or material compositions provide high freedom in designing spatial inhomogeneity over an optically thin interface. A number of planar optics such as filters\cite{cheng2015structural,proust2016all}, lenses\cite{khorasaninejad2016metalenses,khorasaninejad2017metalenses}, polarizers\cite{yu2012broadband,guo2016broadband} and absorbers\cite{liu2017experimental,azad2016metasurface} have been enabled by multifarious reflective or transmissive metasurfaces, featuring high optical performance as well as compact structures. Two central problems arise in the designing process of the metasurfaces. The first is to obtain an accurate prediction of the optical spectrum given a structure, named "Forward Simulation". Traditional methods settle this problem by solving Maxwell's equations approximately using advanced calculations, including rigorous coupled wave analysis (RCWA), finite-difference time-domain method (FDTD), finite-element modeling (FEM) method and so on. The second is to find an optimal structure based on actual demands,  where a nanostructure is generated when given a desired optical response as input, named "Inverse Design". Inverse design of photonic structures were conventionally demonstrated using adjoint sensitivity analysis\cite{piggott2015inverse,piggott2017fabrication,frandsen2016inverse,cao2003adjoint,phan2019high,molesky2018inverse}. However, effective methods are still time-consuming and lack generality in most cases, people have to find a near-optimal solution through a traversal in a limited database, which contains a finite parameter space and corresponding spectrums generated by forward simulation.

Deep learning allows computational models that are composed of multiple processing layers to learn representations of data with multiple levels of abstraction. These methods have dramatically improved the state-of-the-art in computer vision, natural language processing, speech recognition ,and other applications\cite{lecun2015deep}. Deep learning has also been successfully applied to conventional science and engineering fields outside of computer science, such as condensed matter\cite{carrasquilla2017machine}, particle physics\cite{baldi2014searching}, chemical syntheses\cite{segler2018planning}, microscopy\cite{chen2016deep} and proteomics\cite{zhou2017pdeep,gessulat2019prosit}. The strong fitting ability of deep neural networks has also caused quite a stir in the optical community, neural networks (NN) can be used to simulate the optical response of a component (Forward Simulation) as well as design a topology given a desired optical response (Inverse Design). 

\subsection*{Related Work}
In recent years, remarkable achievements with deep learning technologies have been made in the inverse design of optical devices \cite{liu2018training,ma2018deep,malkiel2018plasmonic,peurifoy2018nanophotonic,asano2018optimization,tahersima2018deep,liu2018generative,an2019novel,jiang2019global,jiang2019free}, as well as several optical implementations of  NN\cite{tait2017neuromorphic,mehrabian2018pcnna,lin2018all}.

Many applications of deep learning use feedforward neural network architectures. To go from one layer to the next, a set of units compute a weighted sum of their inputs from the previous layer and pass the result through a non-linear function\cite{lecun2015deep}. The fully connected layer, convolutional layer and transpose convolutional layer are the most basic components of feedforward neural network. Fully connected network (FC) is composed of several fully connected layers, which is the simplest neural network structure and capable of handling one-dimensional vectors. Convolutional neural network (CNN) mainly consists of convolutional layers, which is often used for feature extraction tasks of multidimensional data. The transpose convolution can be considered as an upsampling process as opposed to the convolution process, and the transpose convolution layer is often embedded in many networks related to the generation pattern, like generative adversarial networks (GAN)\cite{goodfellow2014generative} and fully convolutional networks (FCN)\cite{long2015fully}.

Optimization problems in the field of optics are often very complex, but the problem can be modeled in a more regular and simple way by using several one-dimensional parameters, whether the problem itself is one-dimensional, two-dimensional or three-dimensional. D. Liu \emph{et al.} used FC to learn non-unique electromagnetic scattering of alternating dielectric thin films with the varying combination of thickness and materials\cite{liu2018training}. They proposed a tandem architecture combining forward simulation and inverse design to overcome the issue of data inconsistency and slow training process, which has become the dominant architecture for solving similar problems. J. Peurifoy \emph{et al.} adopted a FC having 4 hidden layers, 100 neurons of each to approximate light scattering of multilayer shell nanoparticles of $S_iO_2$ and $T_iO_2$\cite{peurifoy2018nanophotonic}. I. Malkiel \emph{et al.} expounded the relationship between the spectral complexity and design feasibly, then provided an FC with dozen huge layers and multiple input entrance. Their method can be applied to direct on-demand engineering of plasmonic structures and metasurfaces\cite{malkiel2018plasmonic}. Tahersima \emph{et al.} built a robust deeper network on the base of FC with a intensity shortcut proposed in deep residual networks (ResNet)\cite{he2016deep} to inverse design integrated photonic devices, beam splitters for example, whose design space is considerably large\cite{tahersima2018deep}. Recently, the approach brought by S. An \emph{et al.} overcomes three key challenges that have limited previous neural-network-based design schemes: input/output vector dimensional mismatch, accurate EM-wave phase prediction, as well as adaptation to 3-D dielectric structures\cite{an2019novel}.

There is still some work focused on multidimensional representations to meet better requirements. T. Asano \emph{et al.} provided a four-layer neural network including a convolutional layer for prediction of the quality factor in two dimensional photonic crystals\cite{asano2018optimization}. Z. Liu \emph{et al.} proposed a GAN and a simulation neural network that efficiently discovers and optimizes unit cell patterns of metasurfaces in response to user-defined, on-demand spectra at the  input\cite{liu2018generative}. W. Ma \emph{et al.} reported a multi-task model, i.e. broke the task down into a primary task and an auxiliary task, comprising two bidirectional neural networks assembled by a partial stacking strategy, to automatically design and optimize three-dimensional chiral metamaterials \cite{ma2018deep}. J. Jiang \emph{et al.} showed that GAN can train from images of periodic, topology-optimized metagratings to produce high-efficiency, topologically complex devices operating over a broad range of deflection angles and wavelengths\cite{jiang2019free}. Their another work transformed a GAN into a global optimizer by replacing the traditional discriminator with adjoint-based optimization algorithm and used gradient estimation method for back propagation, then combined both as a physics-driven, data-free neural network\cite{jiang2019global}.

\begin{figure*}[bt]
	\centering
	\subfloat[]{
		\begin{minipage}{0.8\linewidth}
			\includegraphics[width=1\linewidth]{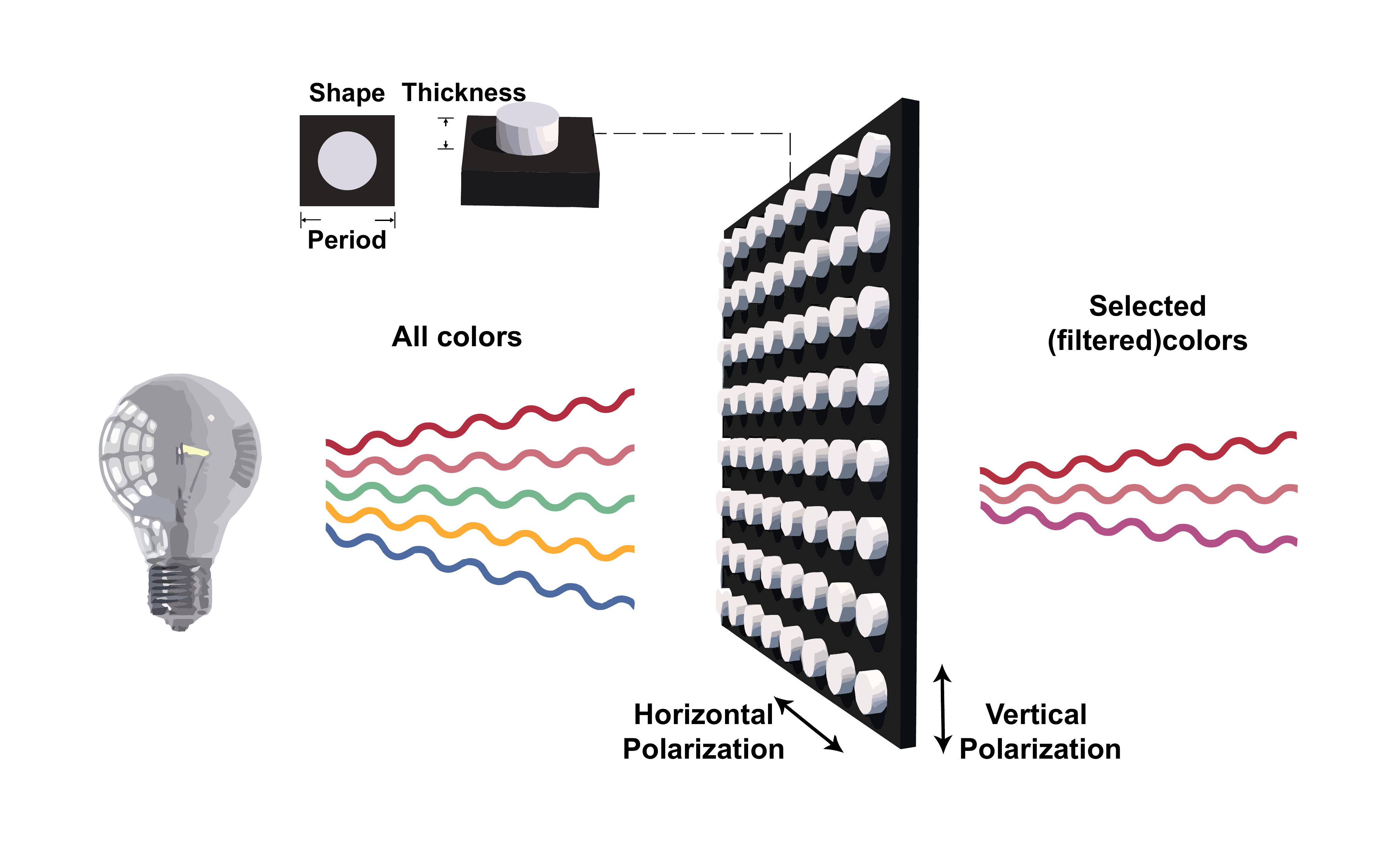}
		\end{minipage}
		\label{first}
	}
	
	\subfloat[]{
		\begin{minipage}{0.48\linewidth}
			\includegraphics[width=1\linewidth]{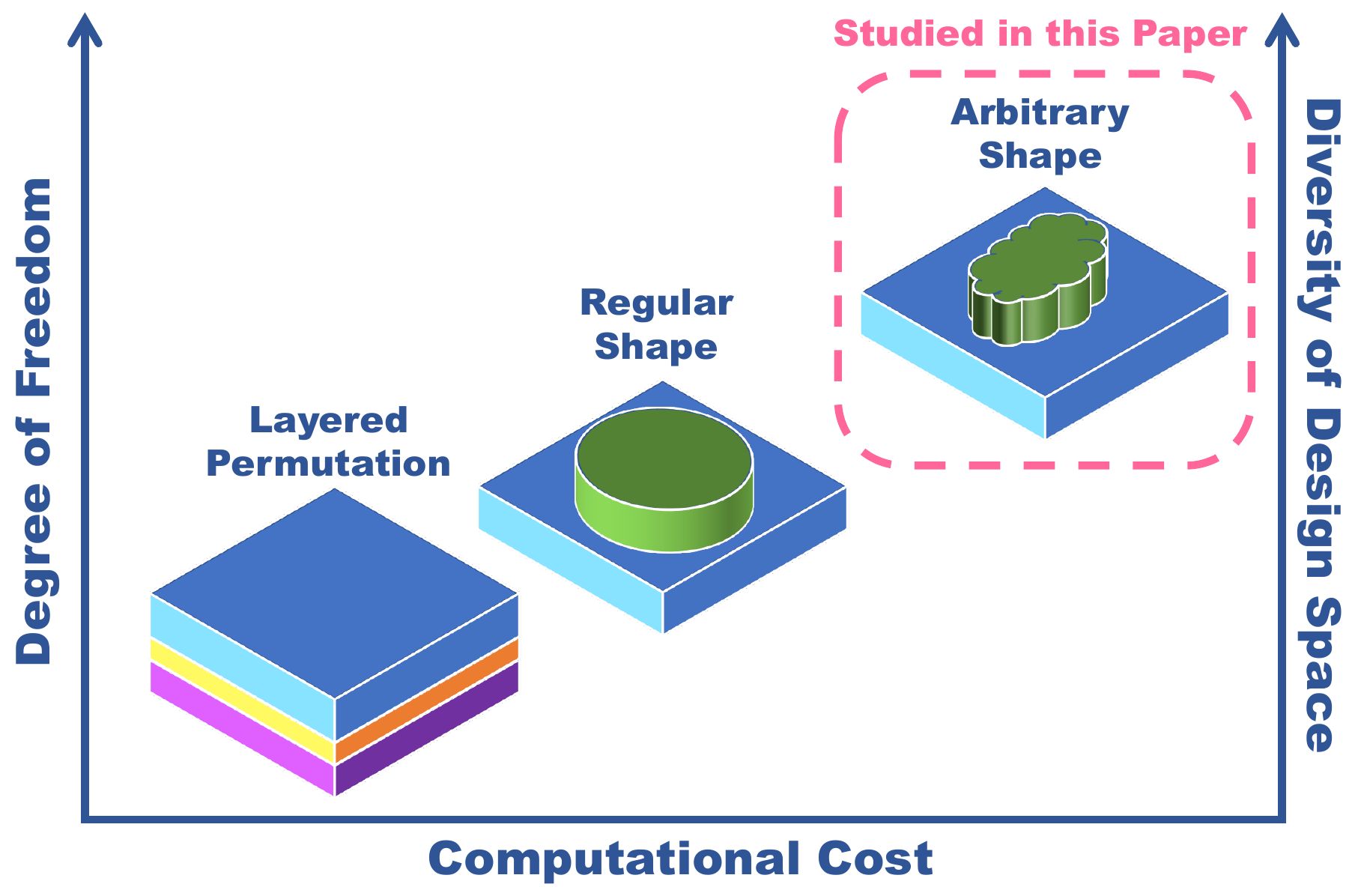}
		\end{minipage}
		\label{cc_dof}
	}
	\subfloat[]{
		\begin{minipage}{0.46\linewidth}
			\includegraphics[width=1\linewidth]{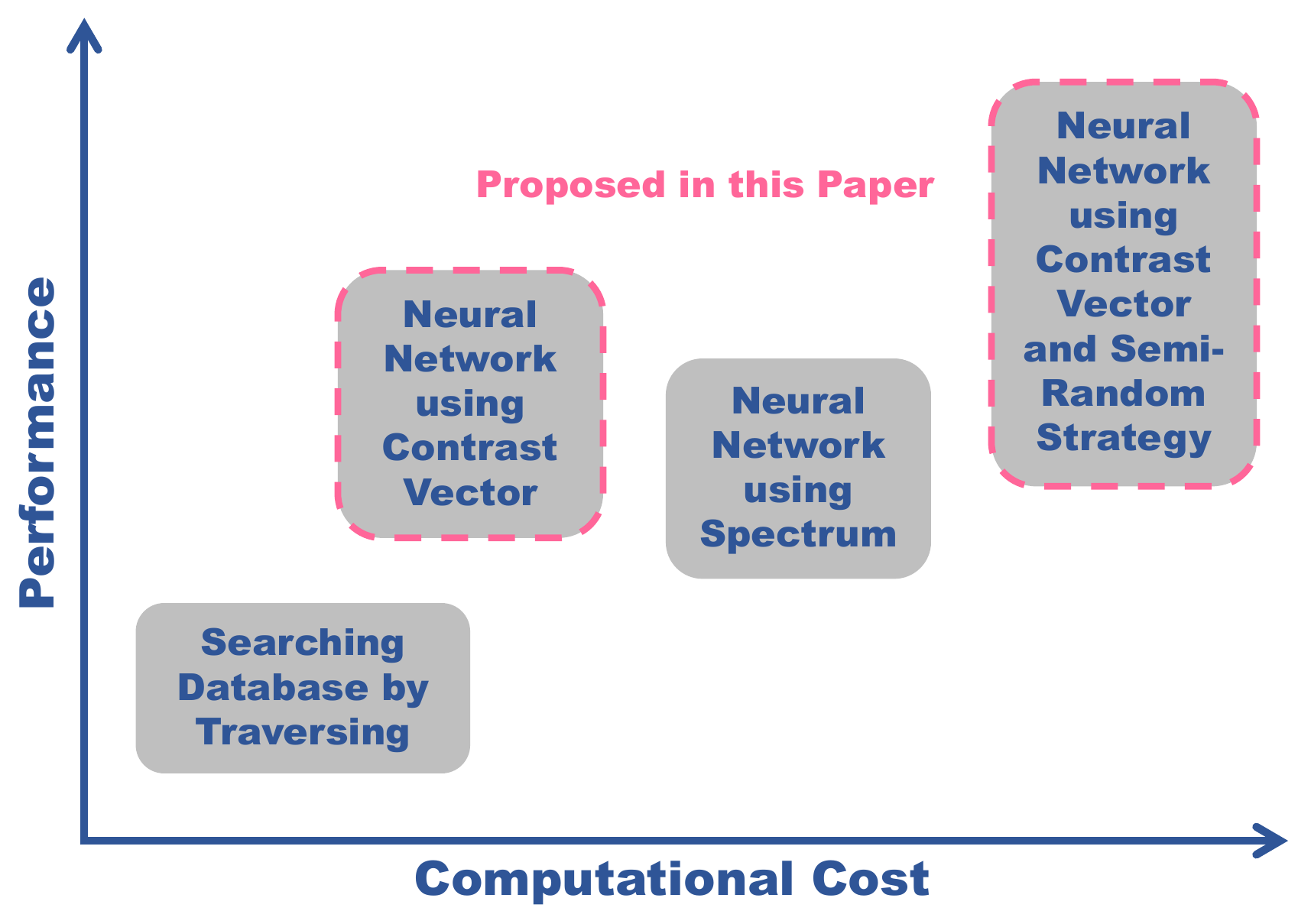}
		\end{minipage}
		\label{cc_per}
	}
	
	\caption{(a) Schematic diagram of the structure to be designed. (b) The tradeoff between the degree of freedom and computational cost. (c) The comparisons between the four methods discussed in this paper. }
\end{figure*}

\subsection*{Our Work}
As shown in Figure \ref{first}, we primarily focus on the filtering function of resonant nanostructures which have been intensively studied, to be specific, the design of color filters based on 2-D periodic grating structures in the visible band. The parameters of devices such as various materials, shapes and layered permutations\cite{liu2018training} determine the degree of freedom. Under the same restriction conditions, a higher degree of freedom granted to the unit cell enhances the probability of generating a qualified spectrum. Enlightened by previous research\cite{liu2018generative,malkiel2018plasmonic,an2019novel}, we improved the degree of freedom a lot because the shape of device is described as an arbitrary binary picture representing the pattern of material on a substrate rather than a regular shape. However, both the requirements on network performance and the computational cost grow at the same time, illustrated in Figure \ref{cc_dof}. Our target is to implement the inverse design with the highest degree of freedom and the least computational cost.

On the grounds of the wide range of wavelength involved in the previous research\cite{malkiel2018plasmonic,an2019novel,liu2018training,liu2018generative}, more suitable spectra for the input constraints can be found to a certain extent. However, we focus on the visible band with a comparatively much narrower range of wavelength here, hence the speed of generating simulation data slows down so that the results of previous studies are prone to miss the sub-optimal solution due to the limited groping ability of the network. Aiming at this difficult problem, we proposed an encoding method called contrast vector to collect the information beneficial for training and two dedicated networks working in series. Moreover, we also proposed a semi-random method to improve performance, which will be discussed later. Qualitative analysis of traditional and our proposed methods can be found in Figure \ref{cc_per}.

To conclude, our simulator can take place of traditional algorithms such as RCWA with more efficiency. In the same give design space, our generator can produce a generated spectrum for a real or artificial desired spectrum. 

\section{Methods}
Our purpose is to acquire an optimized structure satisfying the specified response in the visible band, which inspires us to expand the search space by increasing the degree of freedom. Generally, a generative model named generator is realized elaborately. Additionally, we adopt a novel contrast vector as well as another neural network named simulator to educate the generator. The generator distills the input spectrum information as guidance to generate a structure satisfying the expected electromagnetic response; the simulator extracts information from the input structure and then gets regression estimation of the spectrum. They two can respectively solve forward simulation and inverse design problem efficiently with small error in the design space for artificial spectral input.

Polycrystalline silicon is chosen as the material of the 2D pattern with a fixed thickness of 500 nm, on a substrate made of silicon dioxide, because of its high refractive index and relatively low loss. Considering prior knowledge and actual manufacture requirements, the period ranges from 200nm to 400nm, and the shape is described by a 64 $\times$ 64 pixelated binary image. Besides, 29 points are used to uniformly quantify a single transmittance spectrum where the wavelength ranges from 400nm to 680nm. These limitations do not affect the universality of our method, which will be discussed later. Apparently, each pair of TE and TM responses at 400-680nm, represented by $\bm{T_{TE}}$ and $\bm{T_{TM}}$ with 29 points respectively, can de spliced into a spectrum $\bm{T}=\{t_1,t_2,\cdots,t_{58}\}$. So the problem can be abstracted as following: given a specified $\bm{T}$, how to use algorithms to generate a structure described by a binary image $\bm{S}$ and a period $P$, whose response is $\hat{\bm{T}}$ that equals to $\mathop{\arg\min}_{\hat{\bm{T}}} Distance(\bm{T},\hat{\bm{T}})$. In other words, it equals to the $\hat{\bm{T}}$ that makes the similarity between  $\bm{T}$ and $\hat{\bm{T}}$ maximum.

To avoid missing the sub-optimal solution due to the limited wavelength range, we use a special encoding method to extract the spectrum information which benefits network training as well. We define the contrast of a certain range in a spectrum as the ratio of the maximum transmittance within this range to its counterpart outside. After that, a contrast vector of a particular spectrum can be obtained by sequentially concatenating several contrasts together. Because we are apt to keep all the values in one range small, we pay special attention to the maximum which represents the change in the spectrum.  For a given $\bm{T_{TE}}$ defined above, let $c_i$ be the value of the $i$th contrast, contrast vector is obtained according to the following algorithm \ref{algo}. Then we have $\bm{C_{TE}}$, and calculate $\bm{C_{TM}}$ identically. Finally, splice $\bm{C_{TE}}$, $\bm{C_{TM}}$ in the same way that spliced $\bm{T_{TE}}$, $\bm{T_{TM}}$ to get $\bm{C}$.

\SetAlgoNoLine
\begin{algorithm}
	\KwData{$\bm{T_{TE}}=\{t_1,t_2,\cdots,t_{29}\}$}
	\KwResult{$\bm{C_{TE}}$}
	$\Gamma = \{1,2,\cdots,7\}$\;
	$\Theta = \{1,2,\cdots,29\}$\;
	\For{$i\in\Gamma$}
	{
		$\Omega=\{4i-3,4i-2,\cdots,4i+1\}$\;
		\For{$k\in \Omega$} 
		{
			$max_{in} = \max t_k$\;
		}
		\For{$l\in \complement _{\Theta} \Omega$}
		{
			$max_{out} = \max t_l$\;
		}
		$c_i=\dfrac{max_{in}}{max_{out}}$
	}
	$\bm{C_{TE}}=\{c_1,c_2,\cdots,c_{7}\}$
	\caption{Calculating contrast vector according to the transmittance spectrum}
	\label{algo}
\end{algorithm}

\begin{figure*}[bt]
	\centering
	\subfloat[]{
		\begin{minipage}{0.33\linewidth}
			\includegraphics[width=1\linewidth]{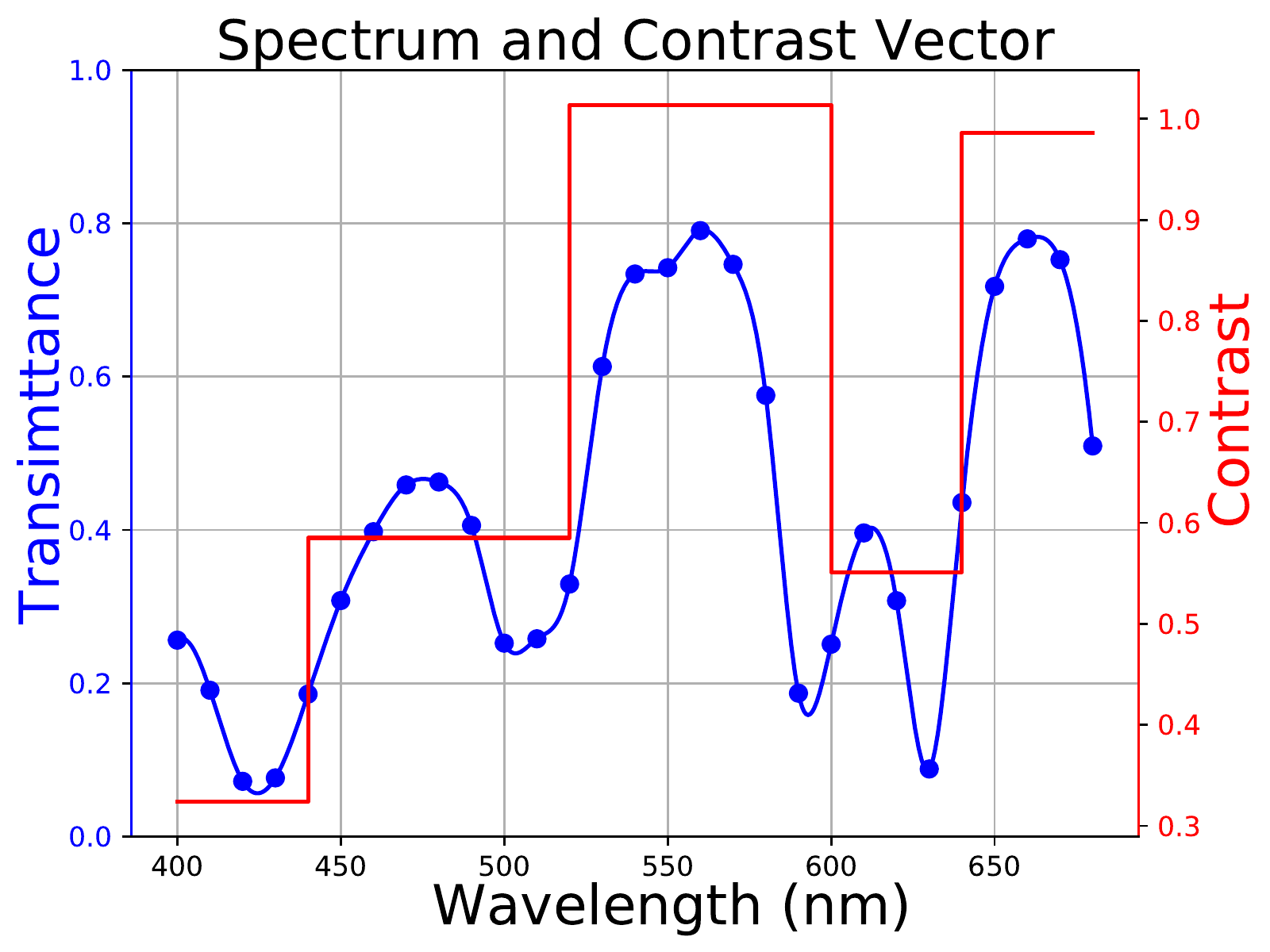}
		\end{minipage}
		\label{CV}
	}
	\subfloat[]{
		\begin{minipage}{0.63\linewidth}
			\includegraphics[width=1\linewidth]{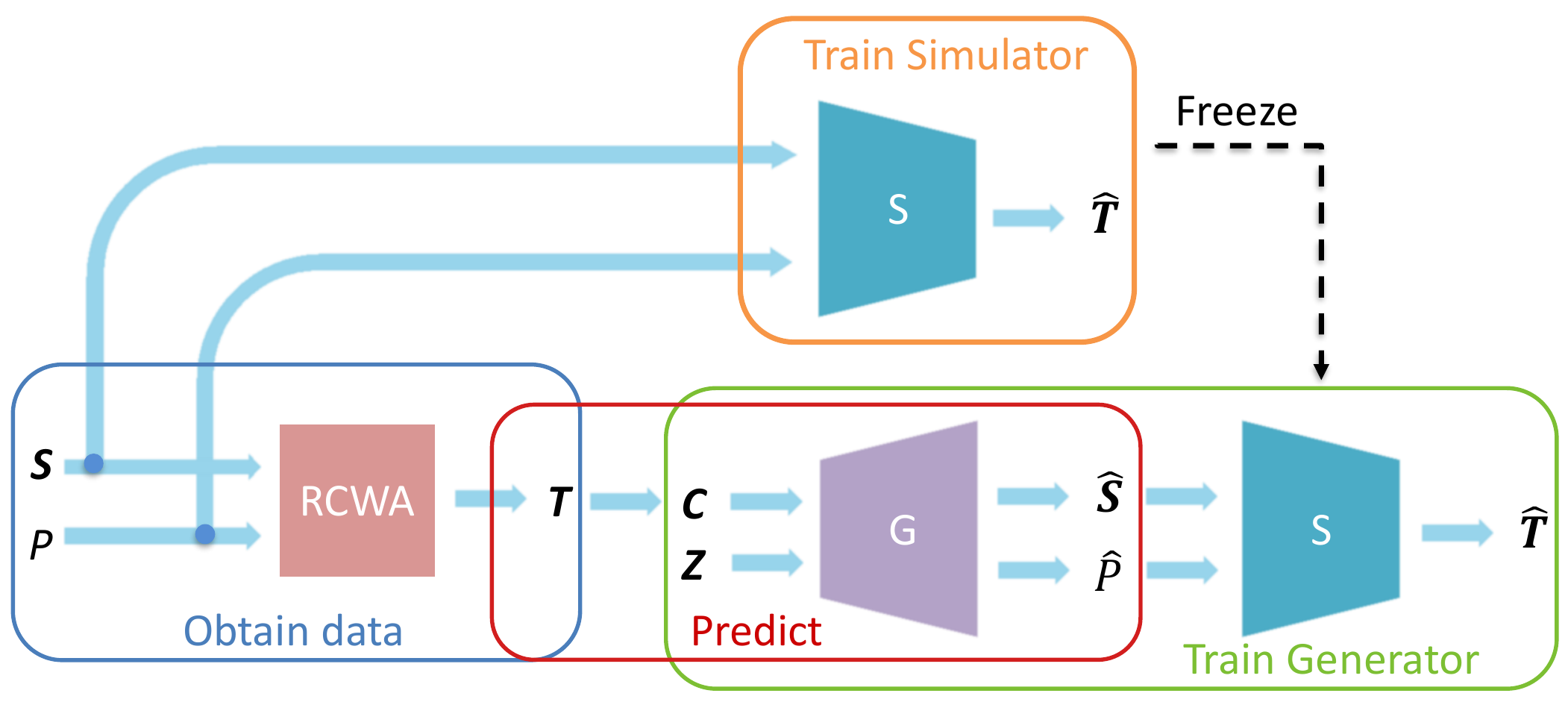}
		\end{minipage}
		\label{pipeline}
	}
	
	\subfloat[]{
		\begin{minipage}{0.7\linewidth}
			\includegraphics[width=1\linewidth]{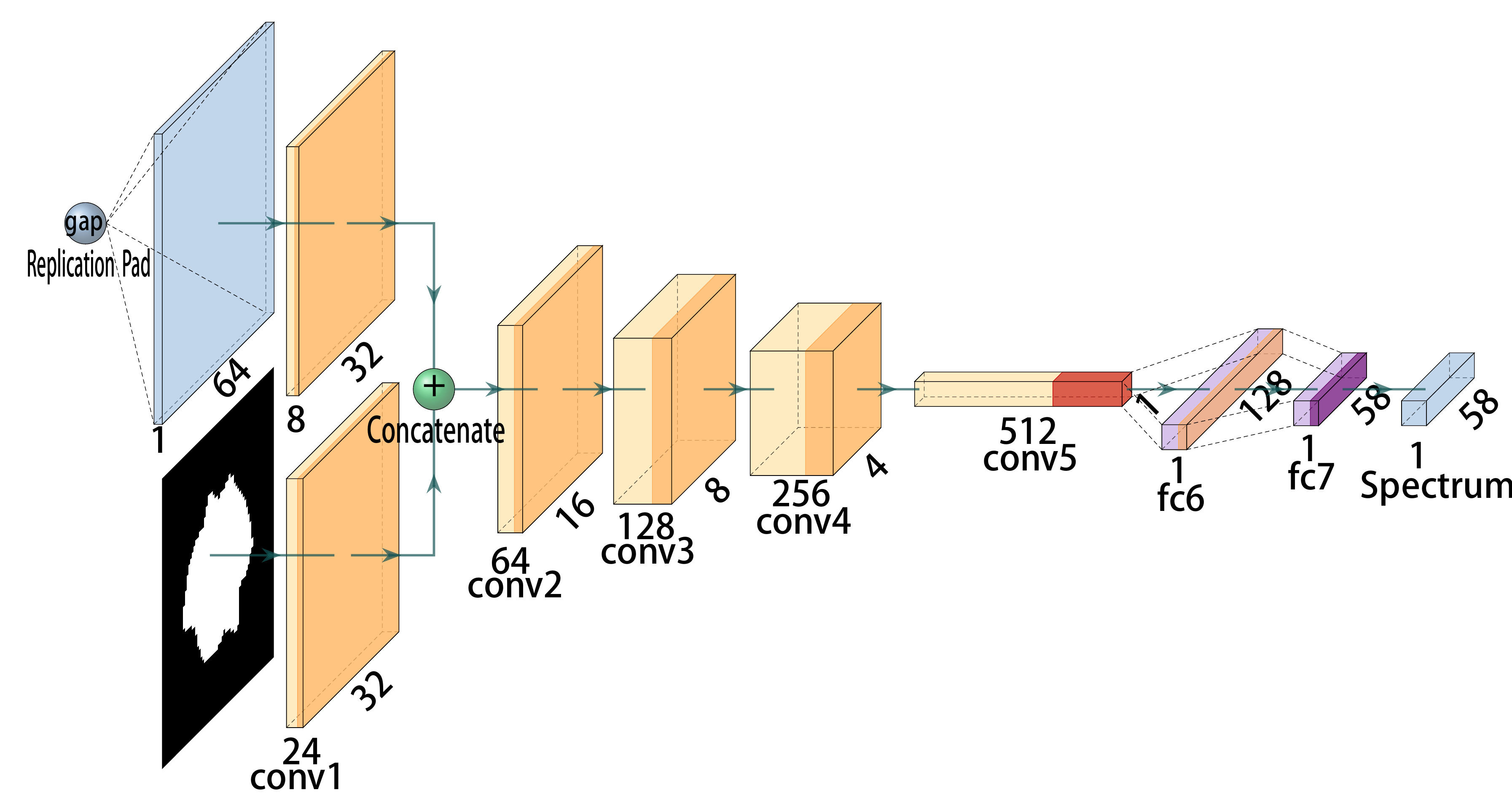}
		\end{minipage}
		\label{simulator}
	}
	
	\subfloat[]{
		\begin{minipage}{0.7\linewidth}
			\includegraphics[width=1\linewidth]{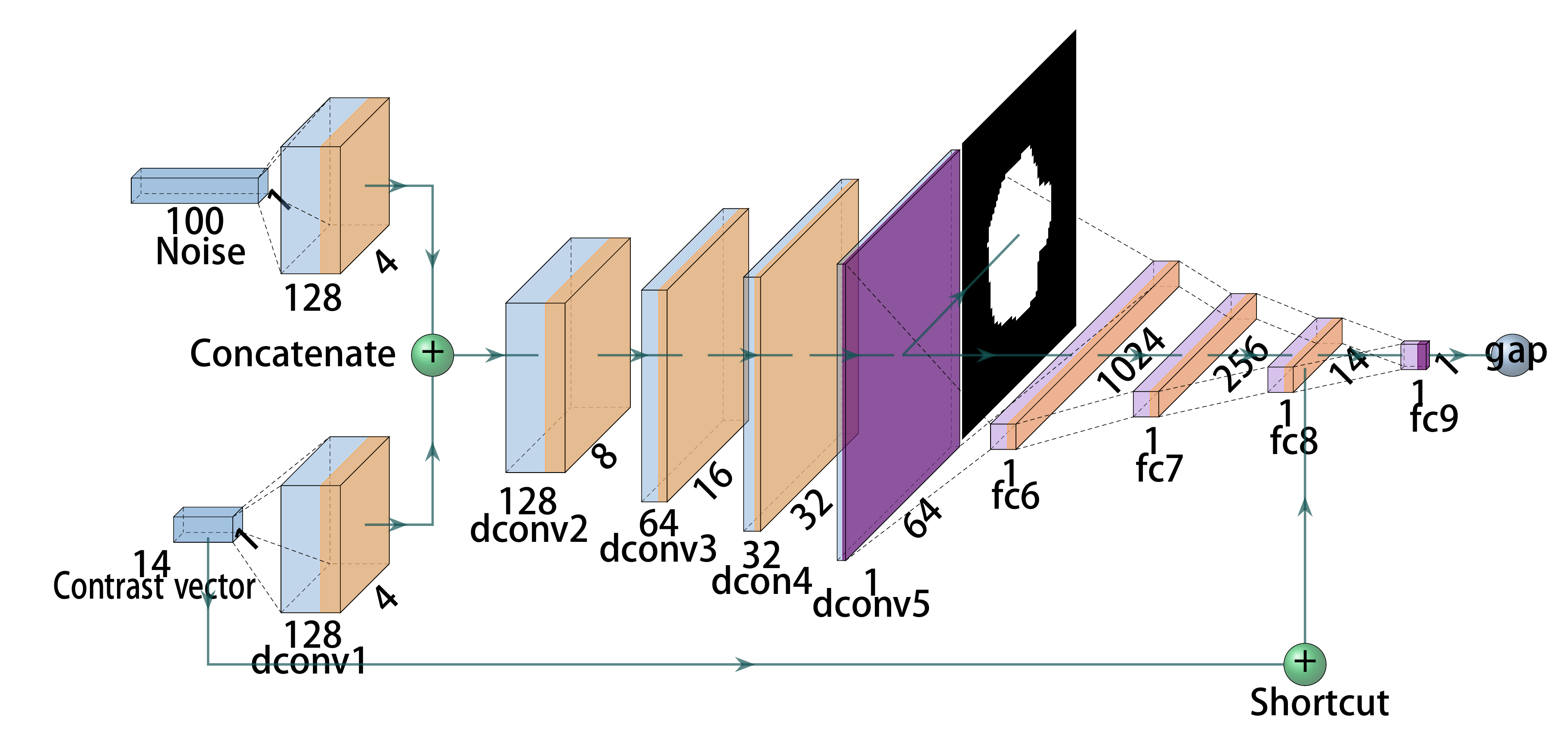}
		\end{minipage}
		\label{generator}
	}
	\caption{(a) Schematic of contrast vector. (b) i.Obtain data using RCWA. ii.Train simulator using these data. iii.Train generator using the same data with frozen simulator. iiii.Use fully trained generator to the inversely design the metasurface filter. (c) The simulator consists of the convolution layers extracting information from images and the fully connected layers converting images into vectors. (d) The generator consists of the deconvolution layers generating images from sequences and the fully connected layers obtaining features from images. For blocks shown in (c)(d), the mapping relationships between color and function are in the Supporting Information \ref{mapping}.}
\end{figure*}

As shown in Figure \ref{CV}, contrast vectors accentuate the peaks and valleys, while ignoring the effect of tiny vibrations in the spectrum. In essence, this weakens the strong correlation between the expected spectrum and the network input in an intelligible way, such that one definite network input can correspond to a great quantity of spectra, thus helping the network understand the meaning of artificial input.

Referring to the network structure of deep convolutional GAN (DCGAN)\cite{radford2015unsupervised}, the conditional inputs of conditional GAN (cGAN)\cite{mirza2014conditional} and the shortcut of residual network (ResNet)\cite{he2016deep}, we design a network using Pytorch\cite{paszke2017automatic} that can turn noise input (random seed for generation) into a qualified unit cell (one 2D binary image with one parameter) according to different guide conditions (target spectrum or target contrast vector). In order to simplify the training process, we use the structural similarity index (SSIM)\cite{wang2004image} to evaluate the similarity between two images instead of a discriminator. As illustrated in Figure \ref{pipeline}, we transform the routine GAN training into a supervised and non-alternating one. Detailed network configurations and training methods are presented in the Supporting Information \ref{appendix}.

The simulator is trained with the data produced by the RCWA\cite{RETICOLO}, in which we accomplish augmentation for better performance. The data contain the corresponding spectra, shapes and periods. The inputs of the simulator are the period $P$ as well as the shape $\bm{S}$, and the output is the predicted spectrum $\hat{\bm{T}}$. The data augmentation method is that we rotate the shapes by 90, 180 and 270 degrees respectively, then exchange TE and TM responses for 90 and 270 degrees respectively. This allows the dataset to be expanded to improve the generalization of the model. We train the simulator to replace the traditional RCWA, because RCWA is relatively slow and breaks the gradient backpropagation. The loss function is the mean square error (MSE) between predicted spectrum $\hat{\bm{T}}$ and real spectrum $\bm{T}$ as bellow, where $N=58$.
\begin{equation}
SimulatorLoss = MSE(\hat{\bm{T}},\bm{T}) = \frac{1}{N}\sum_{i=1}^{N}{(\hat{t_i} - t_i)^2}
\end{equation}

The generator produces the patterns based on the desired spectrum. The input of the generator is a contrast vector $\bm{C}$ transformed from the desired spectrum and a random noise $\bm{Z}$. The outputs are the generated shape $\hat{\bm{S}}$ and the period $\hat{P}$. It is worth noting that $\hat{\bm{S}}$ must be refined into a binary image in the process of predicting because our RCWA algorithm only accepts 0 or 1 to specify a material, which means air or polycrystalline silicon. The shape $\bm{S}$ and period $P$ are the inputs of the simulator, which can predict the spectrum and guide the generator correctly. Data augmentation is not applied in the training process of the generator, because it makes the generator converge worse and the training period can be faster.\cite{liu2018training}. Spectrum loss is the MSE between the real spectrum $\bm{T}$ from training data and the simulated spectrum $\hat{\bm{T}}$ from simulator; shape loss is the SSIM between the real shape $\bm{S}$ and the generated shape $\hat{\bm{S}}$; period loss is the MSE between the real period $P$ and the generated period $\hat{P}$. The generator loss comprises these three parts, as shown below. $\alpha$ and $\beta$ are the two hyperparameters describing the relative importance of the three, which can be fine-tuned accordingly.
\begin{footnotesize}
	\begin{equation}
	\begin{aligned}
	&GeneratorLoss \\ &= SpectrumLoss + \alpha\cdot ShapeLoss + \beta\cdot PeriodLoss \\
	&= \frac{1}{N}\sum_{i=1}^{N}{(\hat{t_i} - t_i)^2} + \alpha\cdot SSIM(\hat{\bm{S}},\bm{S}) + \beta\cdot (\hat{P} - P)^2
	\end{aligned}
	\end{equation}
\end{footnotesize}

High-freedom inverse design is hard to achieve with merely a 2-D array or picture denoting the device since the information is limited. Therefore, several other essential parameters of the metasurface are considered as well in our method. In order to assist the simulator in feature extraction while underlining the importance of periods, the shape along with the period are supplied to the simulator at the same time. However, if we feed the shape as a 2-D array and the period as a real number, the simulator architecture will be asymmetric, leading to an impractical convolutional operation. Thus, the period is duplicated and expanded to a 2-D array in the same size as the shape, then concatenated with the shape to constitute multiple channels for the convenience of simultaneous convolution later. On the other hand, to help the generator extract higher-dimensional features, the noise vector, as well as the contrast vector, are expanded and concatenated. Inside the generator, the shape is produced by transposed convolution layers first, after which the period is generated by fully connected layers. We do not design a generator to get the shape and period with a single network module at the same time, because obtaining these two properties of the metasurface are the two different tasks that cannot use the same network structure with the same weights.

Our network provides a high degree of freedom to the structure to be optimized. It is more efficient than the traditional traversal method when we search the global design space. Because the initial network can map the random noise $\bm{Z}$ into the full design space, the probability of finding the optimal solution is enhanced. For our metasurface filter, the 1-D parameter considered is the period of the unit cell, but it can generally involve any combination of design parameters in the design problems including structure thickness, refractive index, or polarization light, etc. Moreover, as long as the database is sufficient, our network structure is suitable for all metasurface design problems that can be represented as 2-D binary graphs with parameters such as period and thickness.

\section{Results and Discussion}
\begin{figure*}[bt]
	\centering
	\subfloat[]{
		\begin{minipage}{0.31\linewidth}
			\includegraphics[width=1\linewidth]{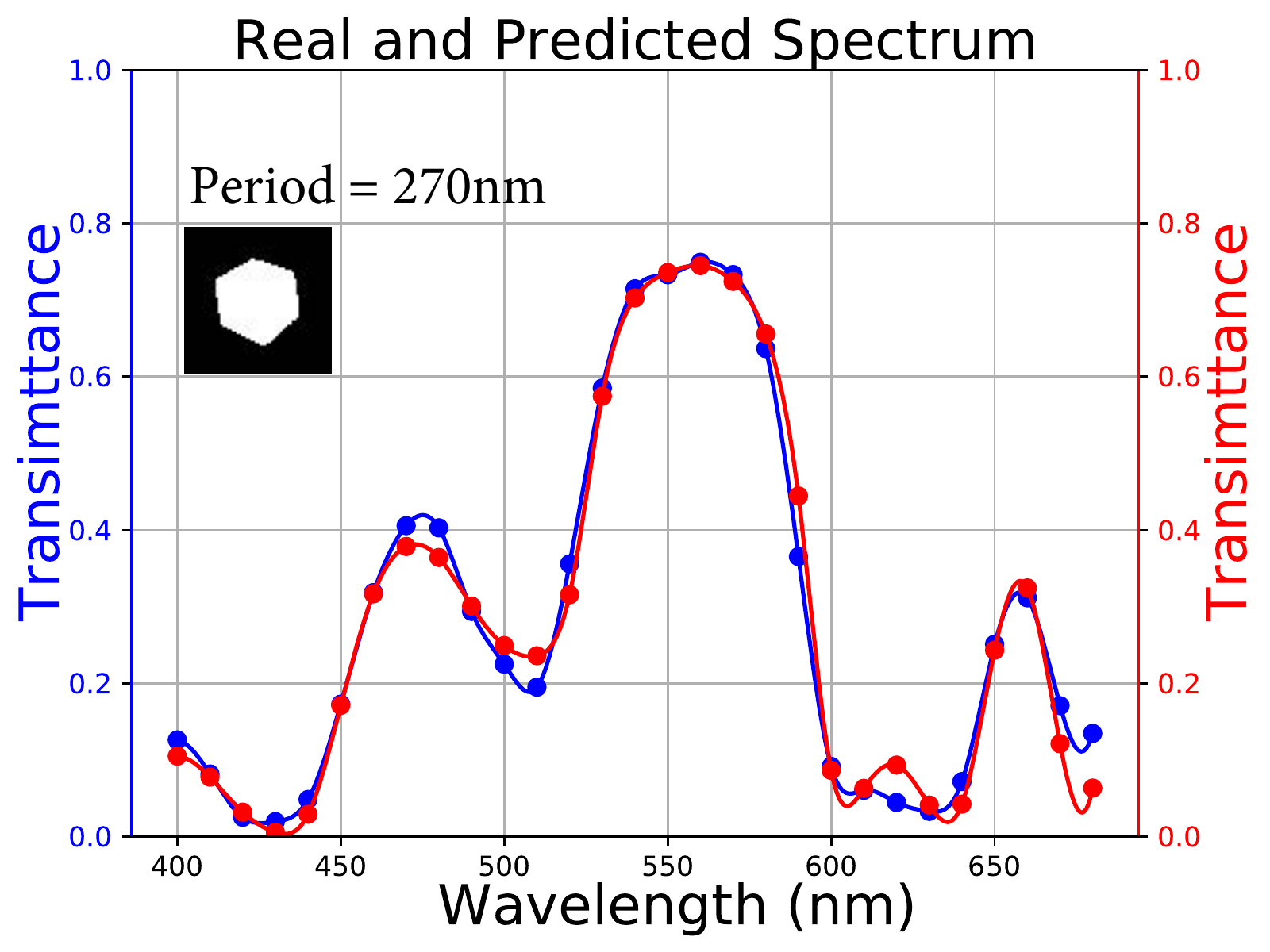}
		\end{minipage}
		\label{t_simulator}
	}
	\subfloat[]{
		\begin{minipage}{0.31\linewidth}
			\includegraphics[width=1\linewidth]{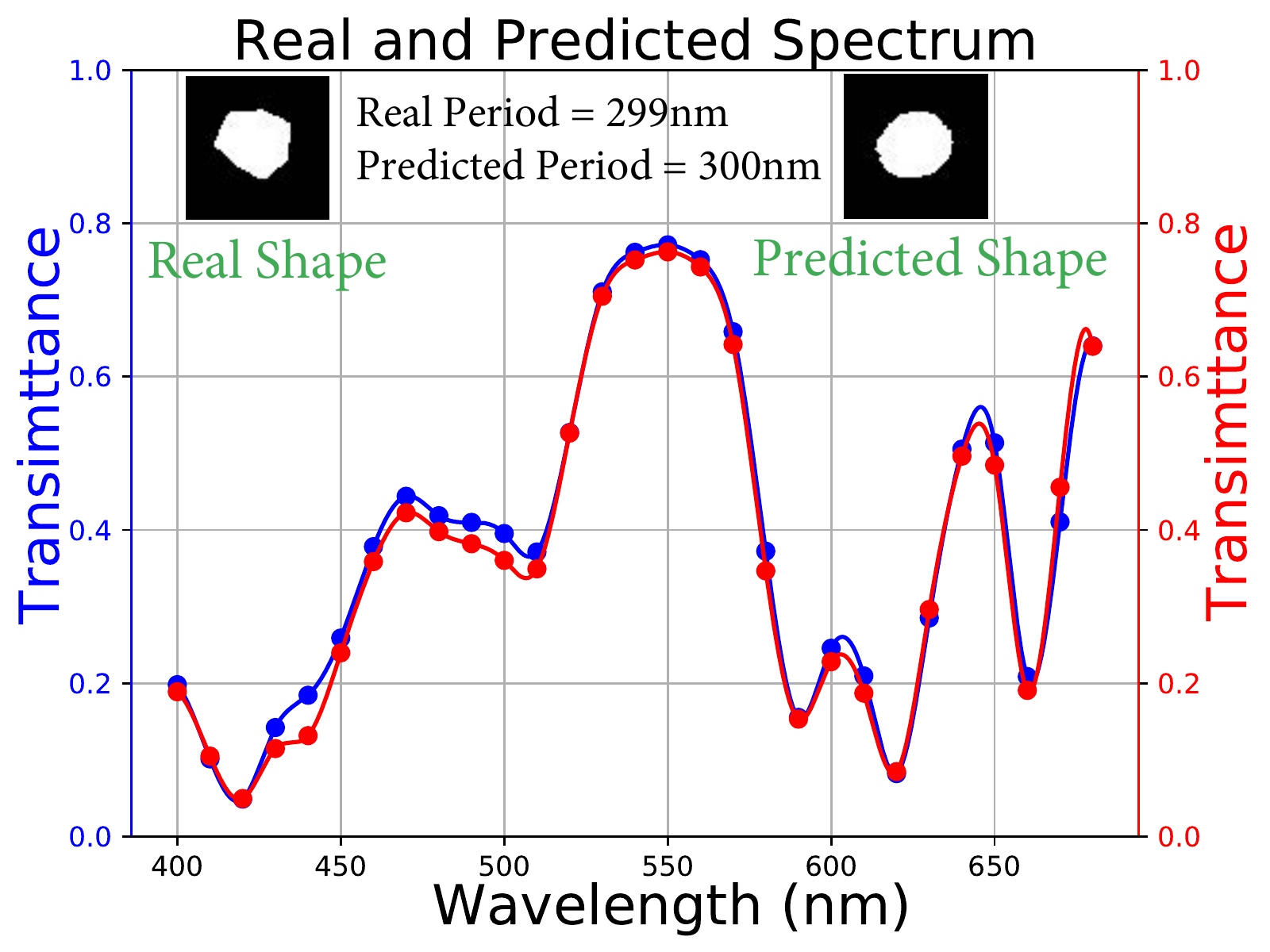}
		\end{minipage}
		\label{t_generator}
	}
	\subfloat[]{
		\begin{minipage}{0.31\linewidth}
			\includegraphics[width=1\linewidth]{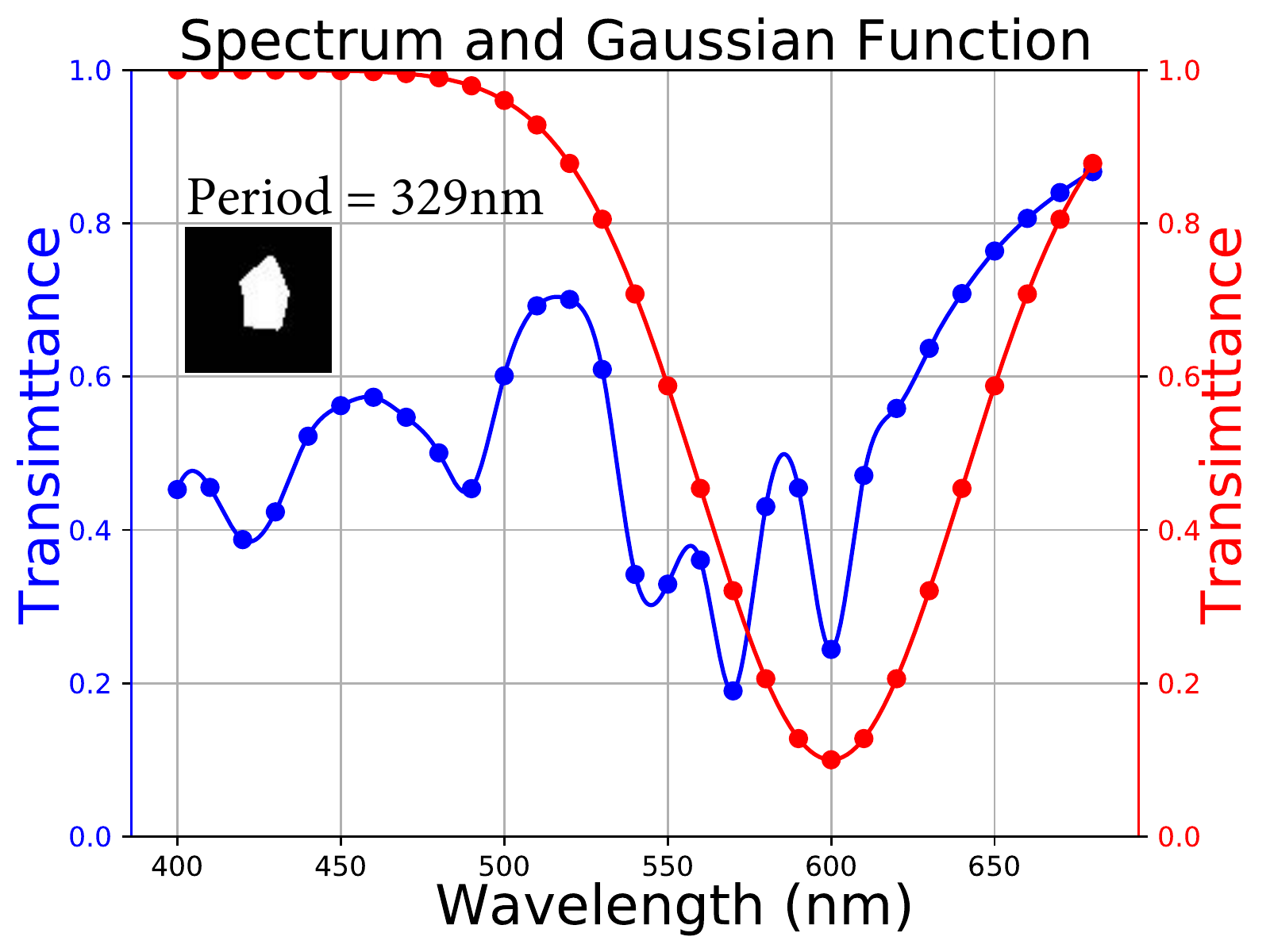}
		\end{minipage}
		\label{c_trad}
	}
	
	\subfloat[]{
		\begin{minipage}{0.31\linewidth}
			\includegraphics[width=1\linewidth]{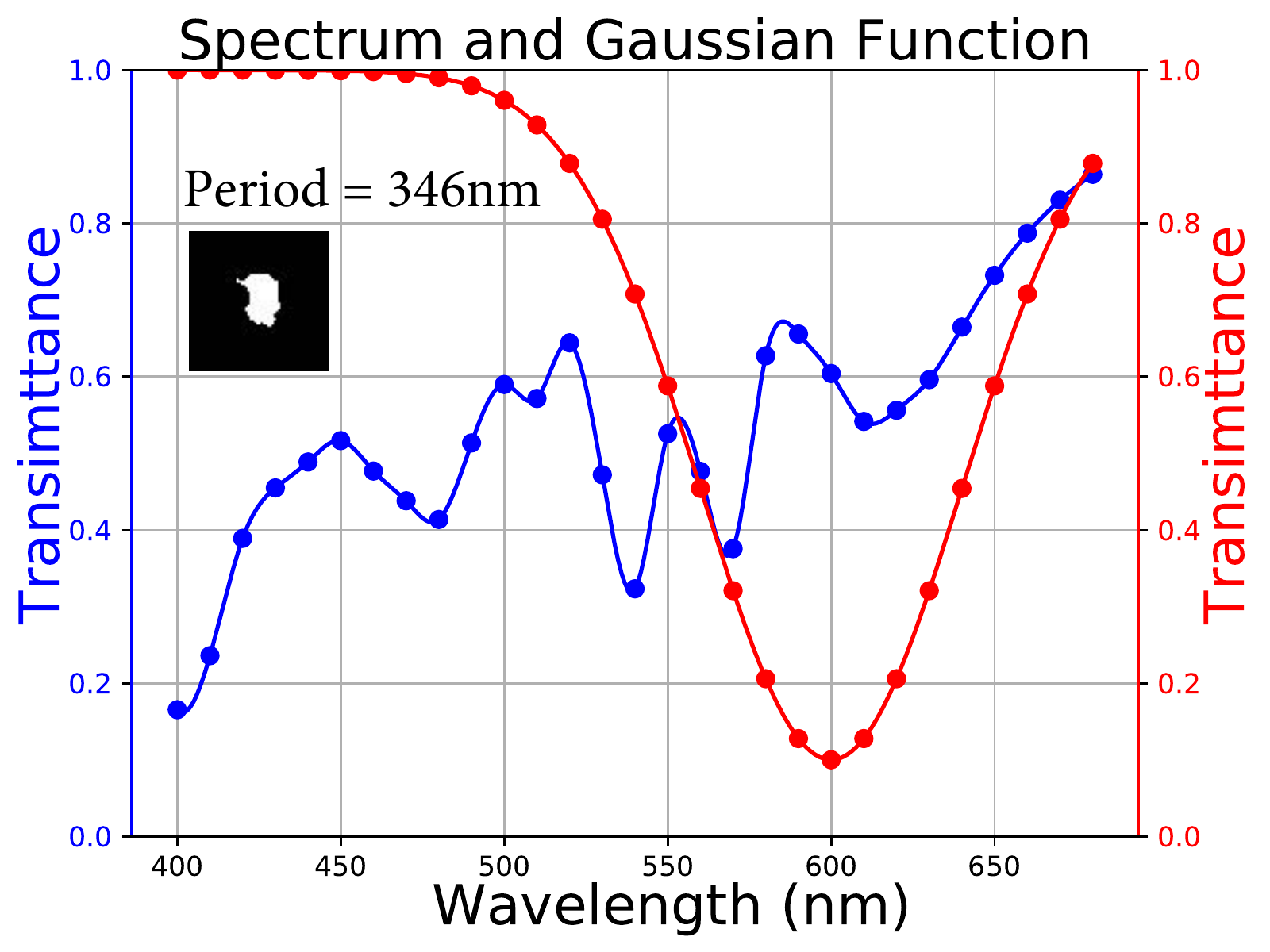}
		\end{minipage}
		\label{c_gauss}
	}
	\subfloat[]{
		\begin{minipage}{0.31\linewidth}
			\includegraphics[width=1\linewidth]{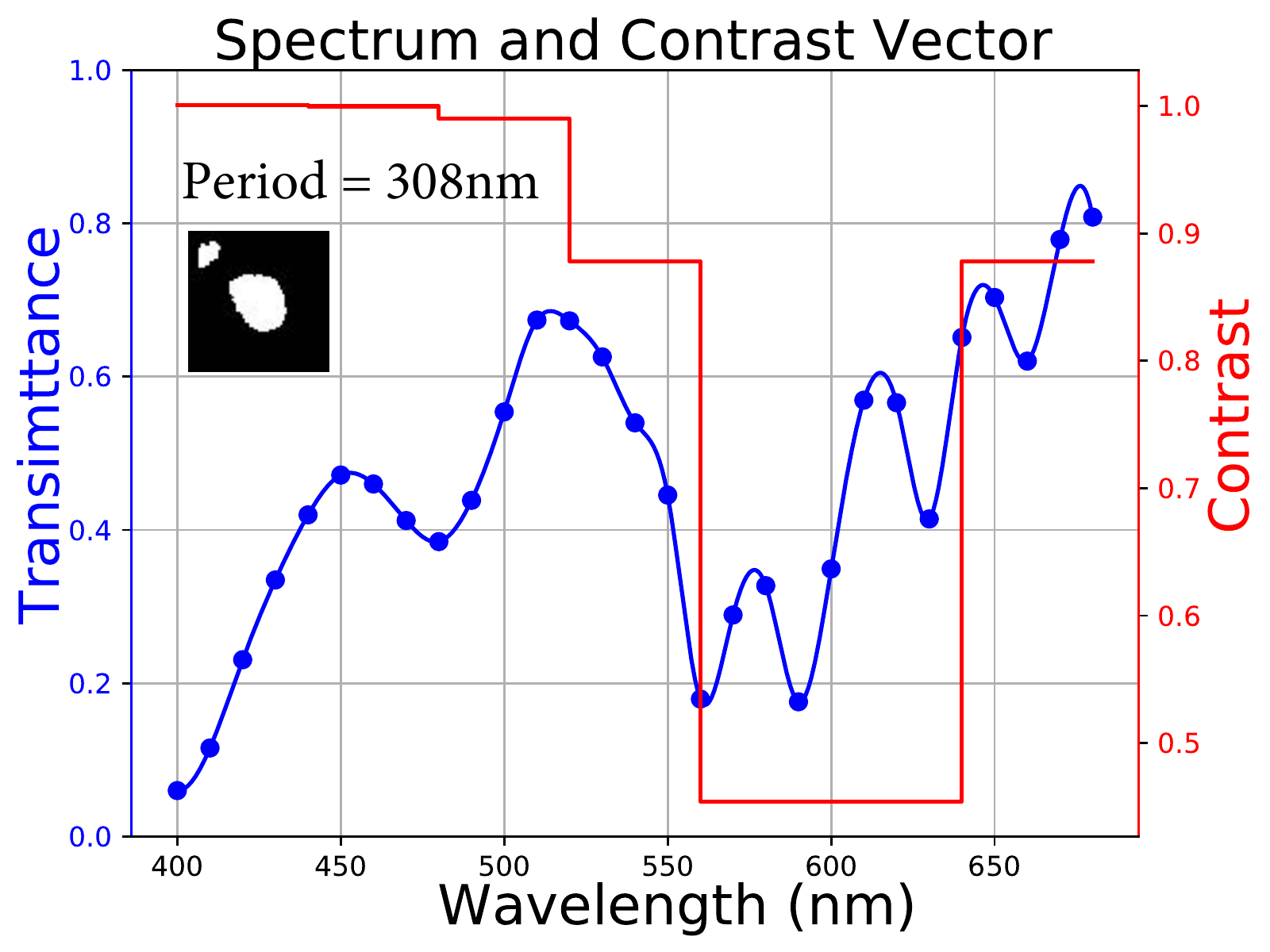}
		\end{minipage}
		\label{c_ctrast}
	}
	\subfloat[]{
		\begin{minipage}{0.31\linewidth}
			\includegraphics[width=1\linewidth]{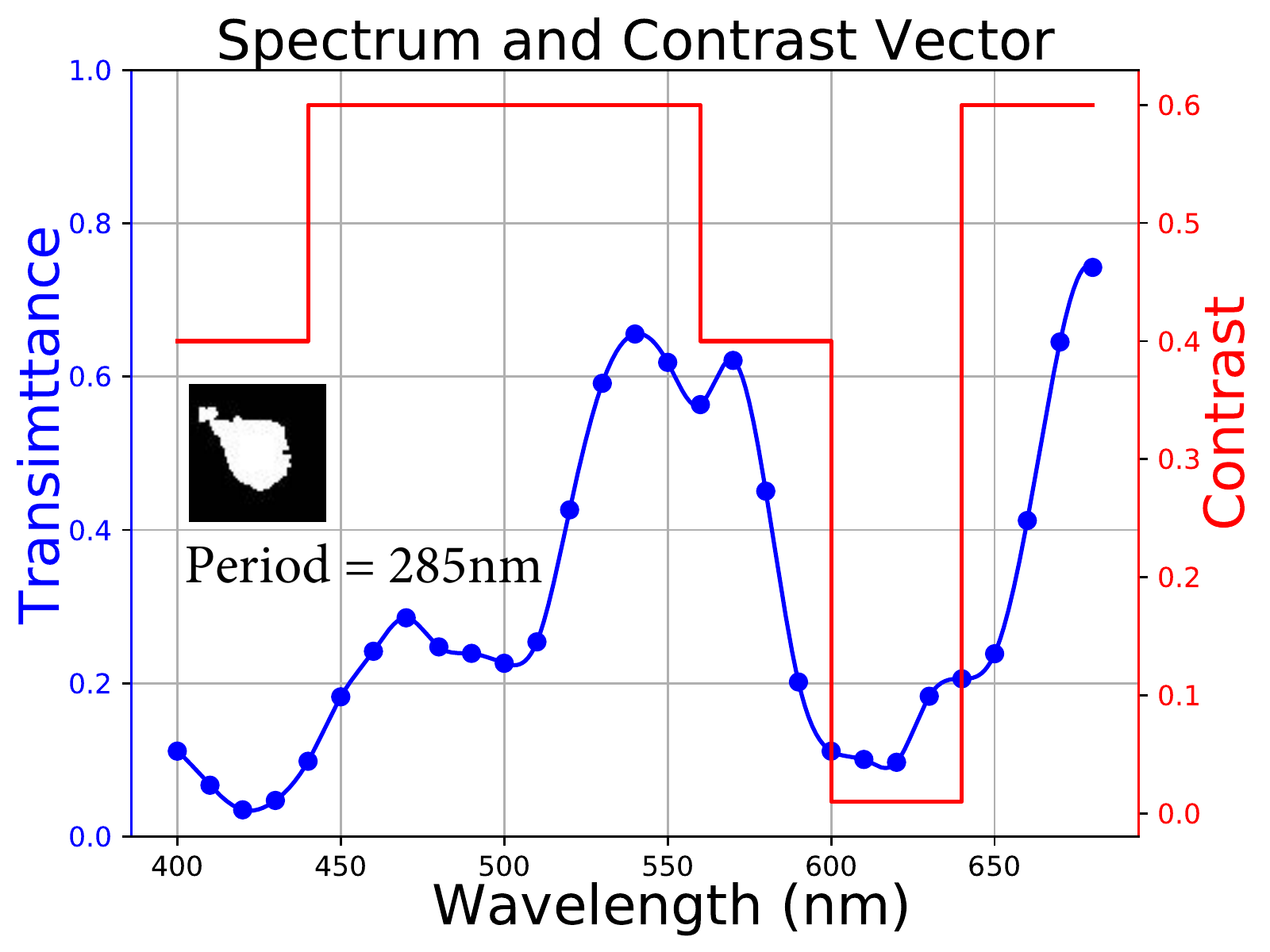}
		\end{minipage}
		\label{c_random}
	}
	\caption{(a) Red curve is a desired spectrum in the validation set. Blue curve is the simulated spectrum from the simulator. (b) Red curve is an desired spectrum in the validation set. Blue curve is the generated spectrum from the generator. (c) Red curve is an artificial Gaussian spectrum. Blue curve is the spectrum with the smallest MSE by traversing in the training set. (d) Red curve is the same artificial Gaussian spectrum with the red curve in c. Blue curve is the generated spectrum from the generator, which is trained without transforming spectra into contrast vectors. (e) Red curve is an artificial group of contrast vectors transformed from the same artificial Gaussian spectrum with the red curve in c. Blue curve is the generated spectrum from the generator. (f) Red curve is an artificial group of contrast vectors whose values are randomly sampled from [0.4,0.5,0.6], while the overall trend is similar with the red curve in e. Blue curve is the generated spectrum from the generator.}
\end{figure*}

To measure the performance of the simulator, we feed it with a randomly generated polygon. The real spectrum and the simulated spectrum are plotted in Figure \ref{t_simulator}. Statistically, the MSE between the real spectrum and the generated spectrum is 4\% to 5\% when we feed 5,200 paired shapes and periods beyond the training set. As for the generator, we feed a contrast vector calculated from the real spectrum into it. The generated and ground truth shape, as well as the period, are shown in Figure \ref{t_generator}. Likewise, the MSE between the desired spectrum and the simulated spectrum of the generated structure is approximately 5\% when the validation set contains 1,300 other real spectra consisting of 58 points each.

\begin{figure*}[bt]
	\centering
	\subfloat[]{
		\begin{minipage}{0.3\linewidth}
			\includegraphics[width=1\linewidth]{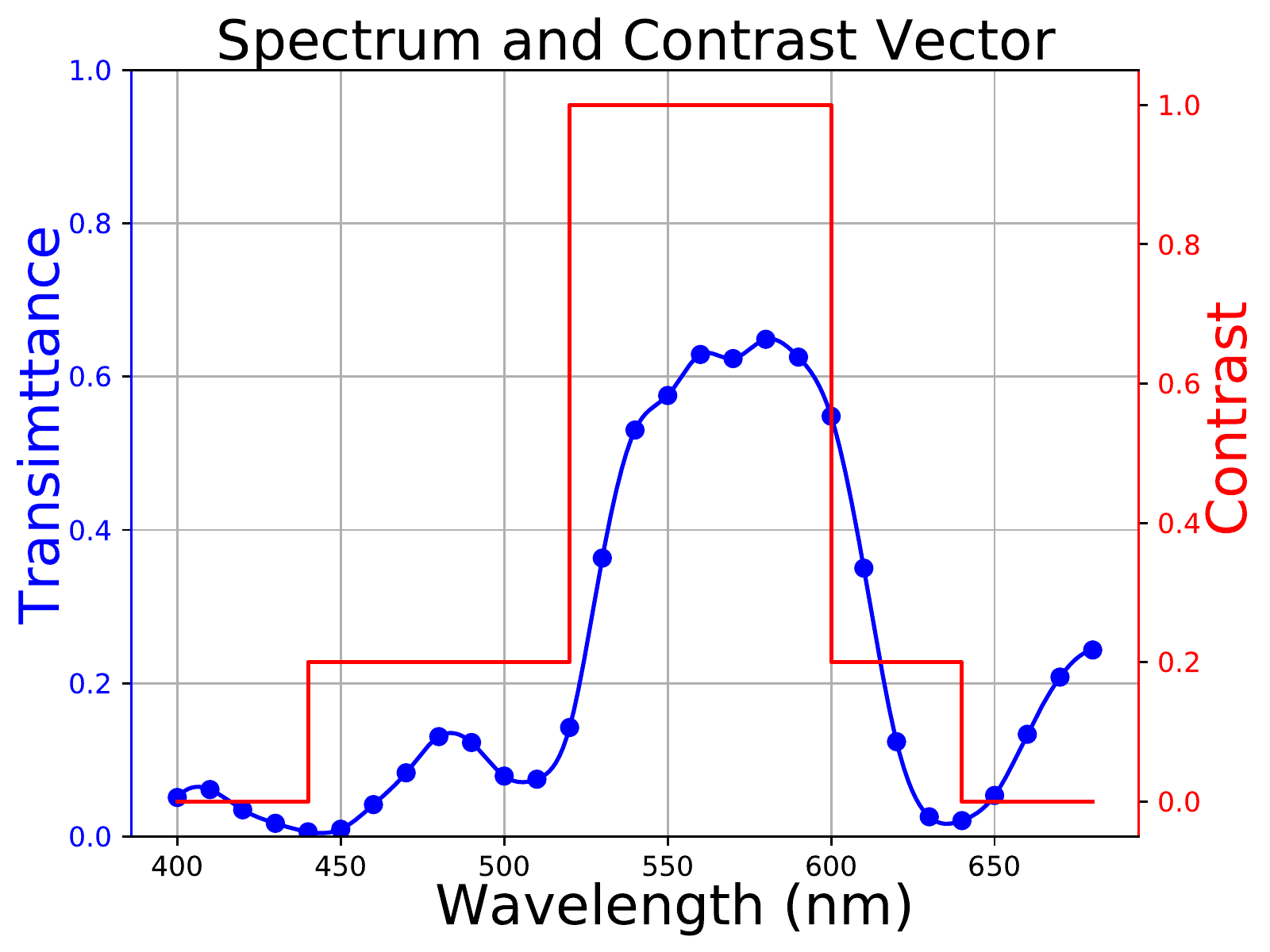}
		\end{minipage}
	}
	\subfloat[]{
		\begin{minipage}{0.3\linewidth}
			\includegraphics[width=1\linewidth]{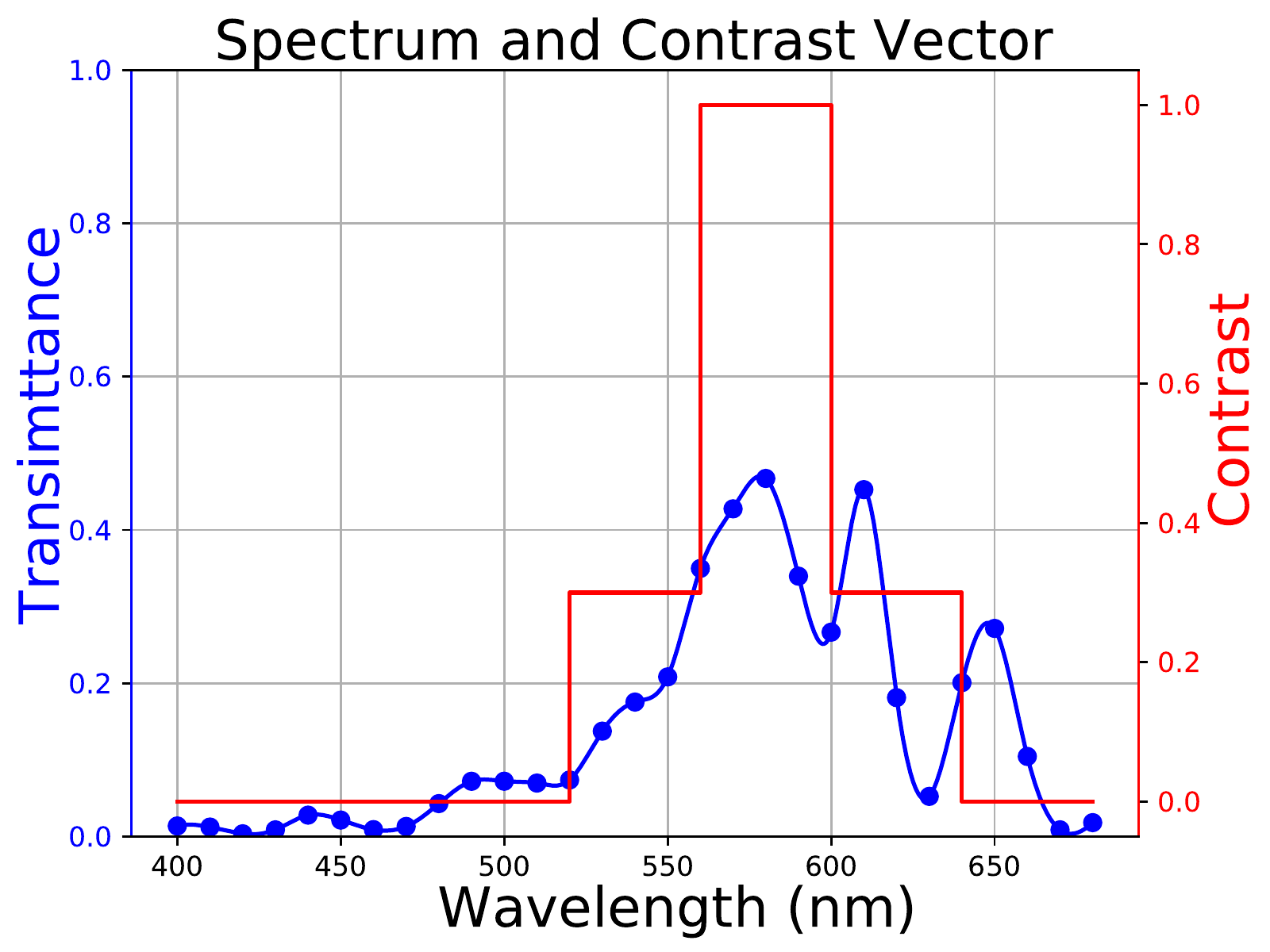}
		\end{minipage}
	}
	\subfloat[]{
		\begin{minipage}{0.3\linewidth}
			\includegraphics[width=1\linewidth]{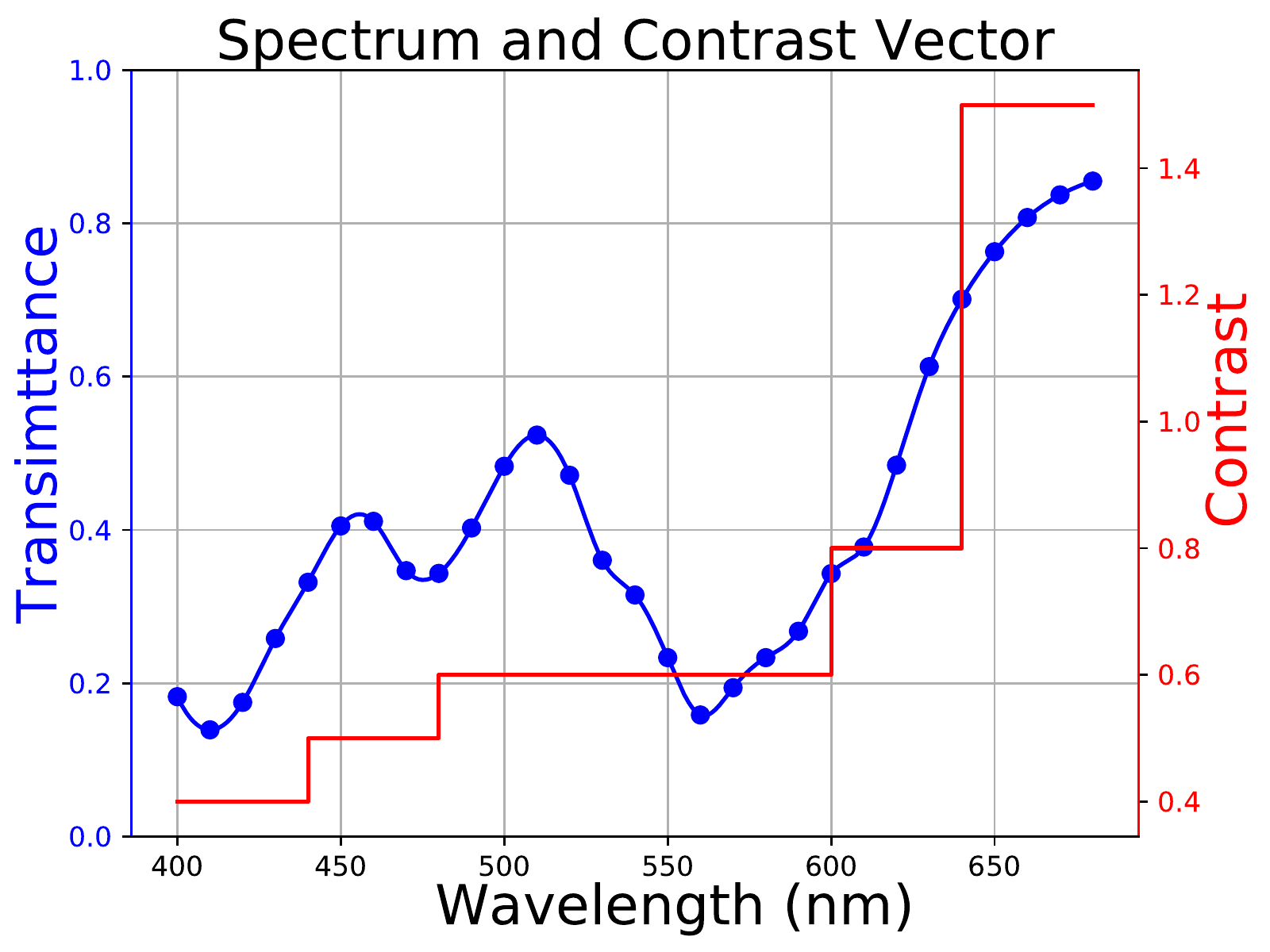}
		\end{minipage}
	}
	
	\subfloat[]{
		\begin{minipage}{0.3\linewidth}
			\includegraphics[width=1\linewidth]{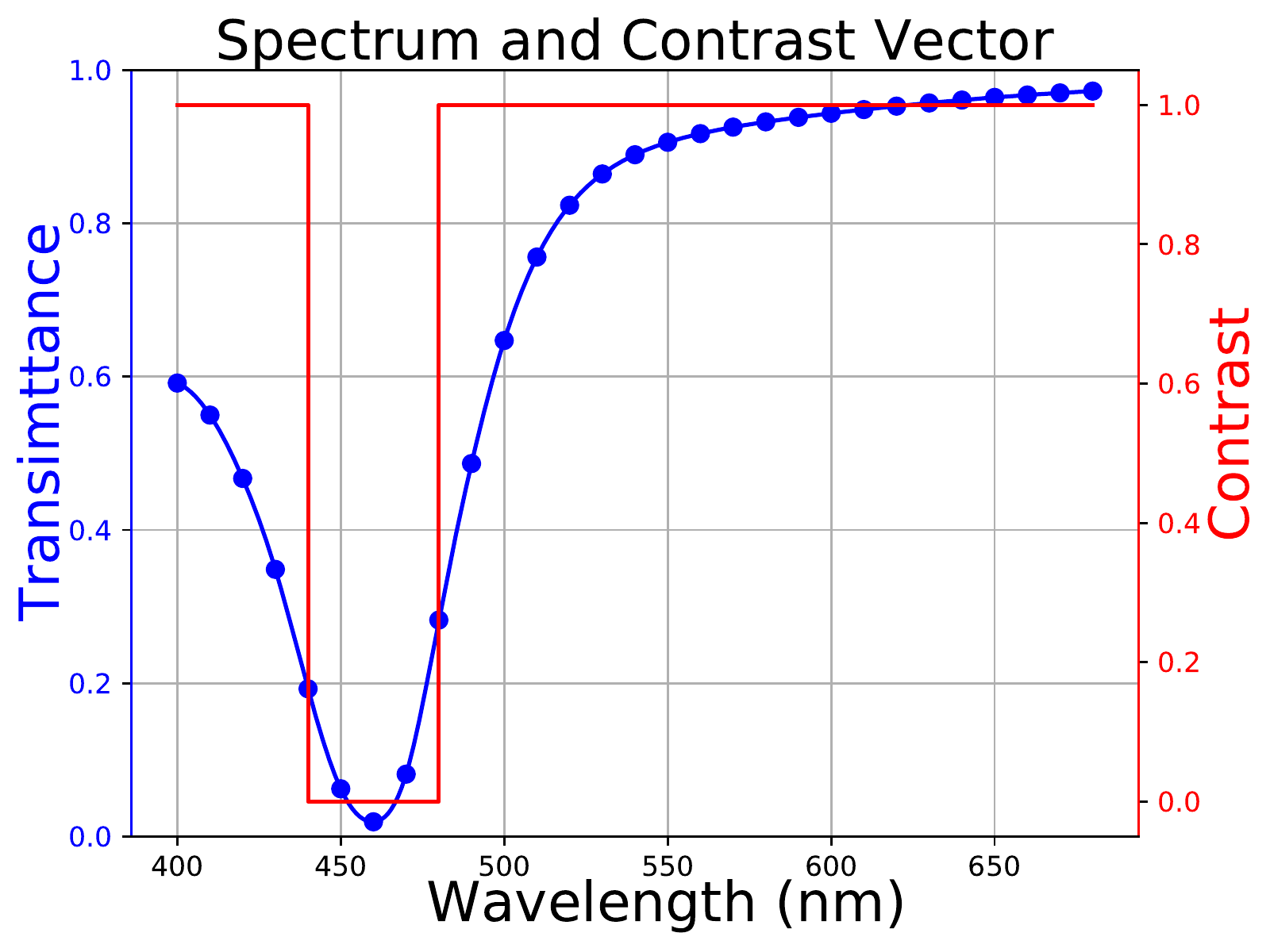}
		\end{minipage}
	}
	\subfloat[]{
		\begin{minipage}{0.3\linewidth}
			\includegraphics[width=1\linewidth]{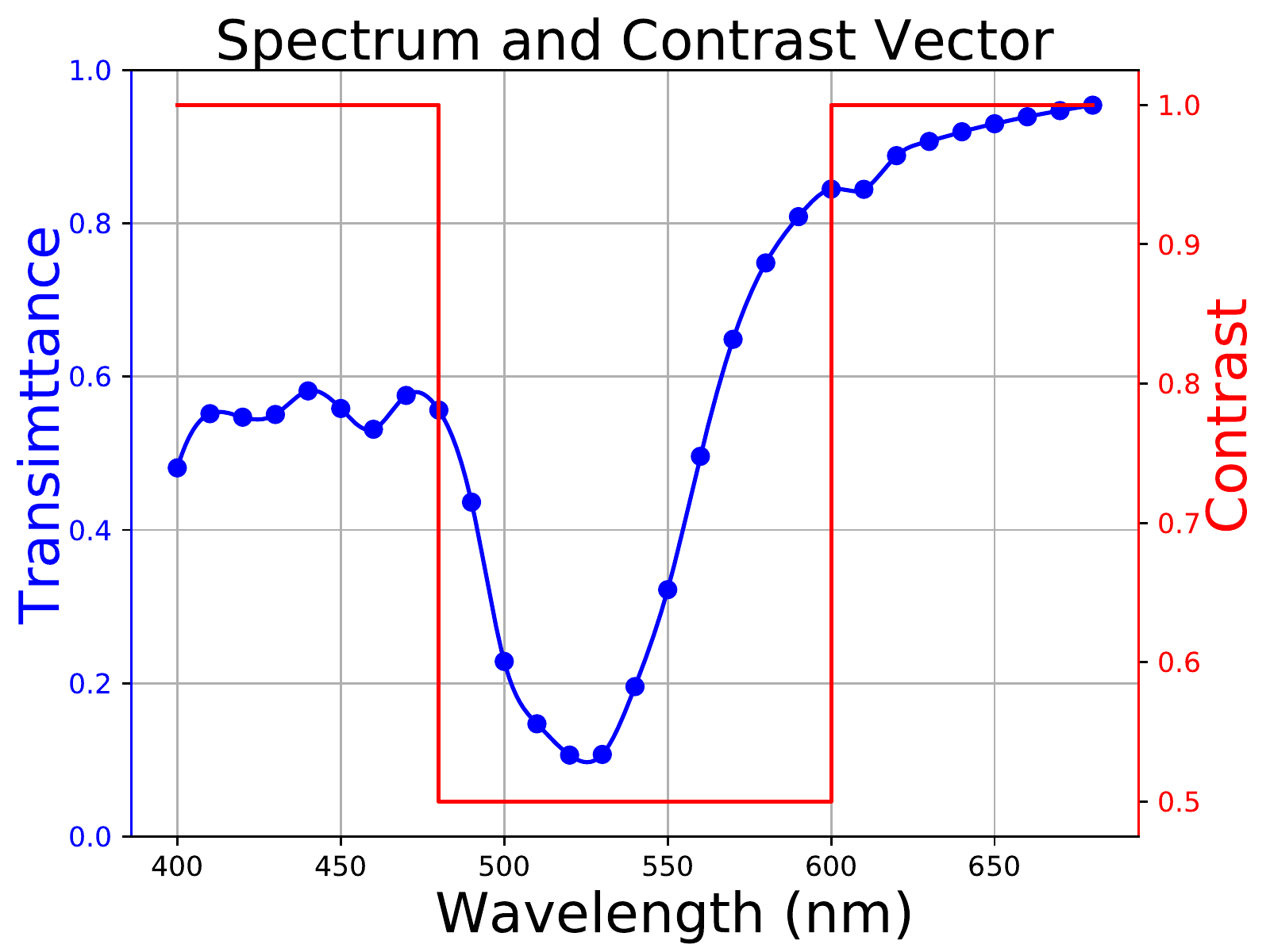}
		\end{minipage}
	}
	\subfloat[]{
		\begin{minipage}{0.3\linewidth}
			\includegraphics[width=1\linewidth]{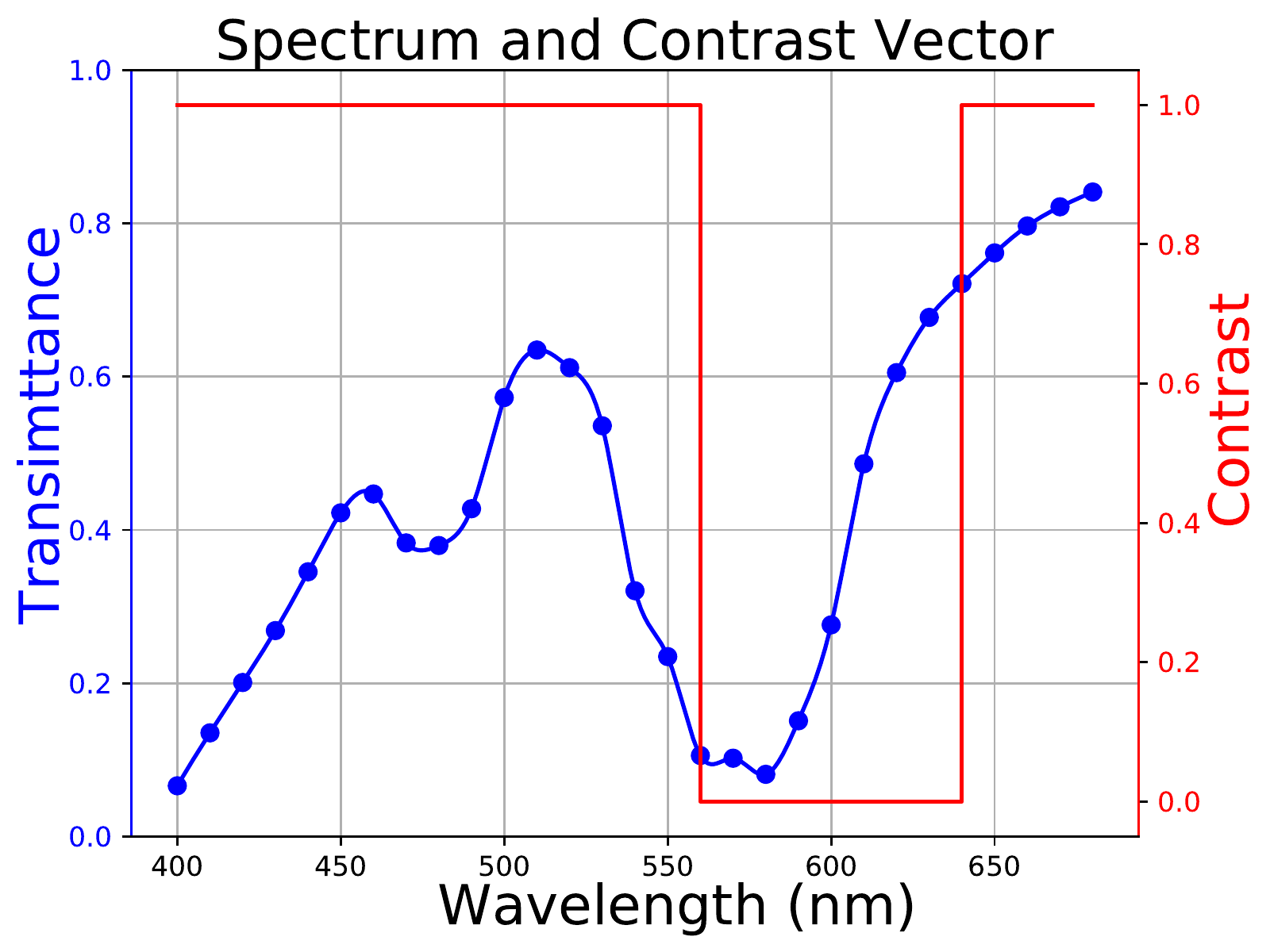}
		\end{minipage}
	}
	\caption{All these test results for generator obtained by using semi-random method mentioned above. (a) Single peak at the green band (492nm - 577nm). (b) Single peak at the yellow band (577nm - 597nm). (c) Single peak at the red band (622nm - 770nm). (d) Single valley at the blue band(455nm - 492nm). (e) Single valley at the green band(492nm - 577nm). (d) Single valley at the yellow band (577nm - 597nm).}
	\label{more}
\end{figure*}

\begin{figure*}[bt]
	\centering
	\begin{minipage}{0.46\linewidth}
		\includegraphics[width=1\linewidth]{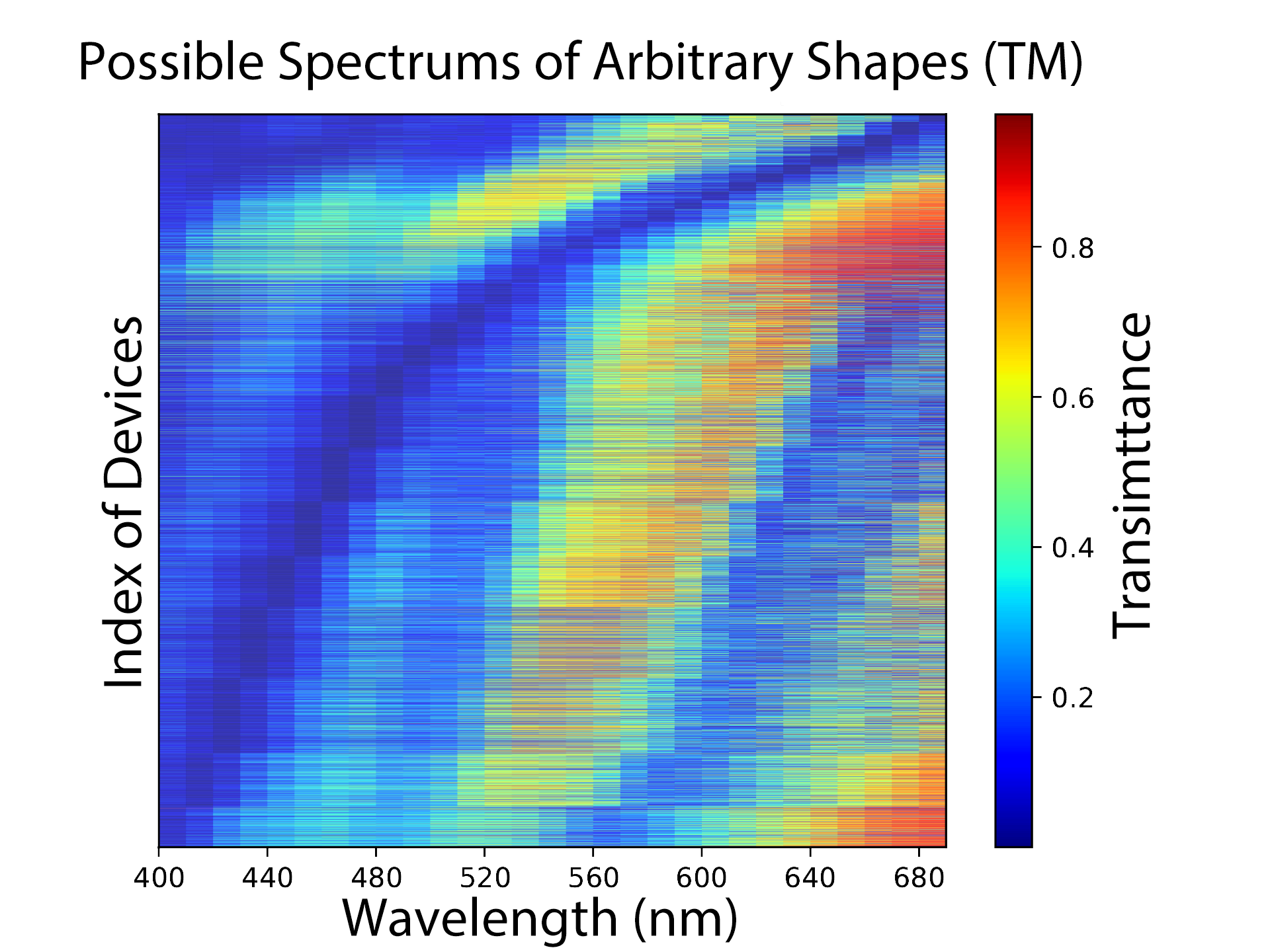}
		\caption{Observations of the spectrum of the entire dataset. Each row represents a real spectrum (only TM is shown because TE is very similar). The sorting rule is that the lower the wavelength corresponding to the minimum transmittance is, the lower the position of this spectrum is.}
		\label{pss}
	\end{minipage}
	\hspace*{1cm}
	\begin{minipage}{0.46\linewidth}
		\includegraphics[width=1\linewidth]{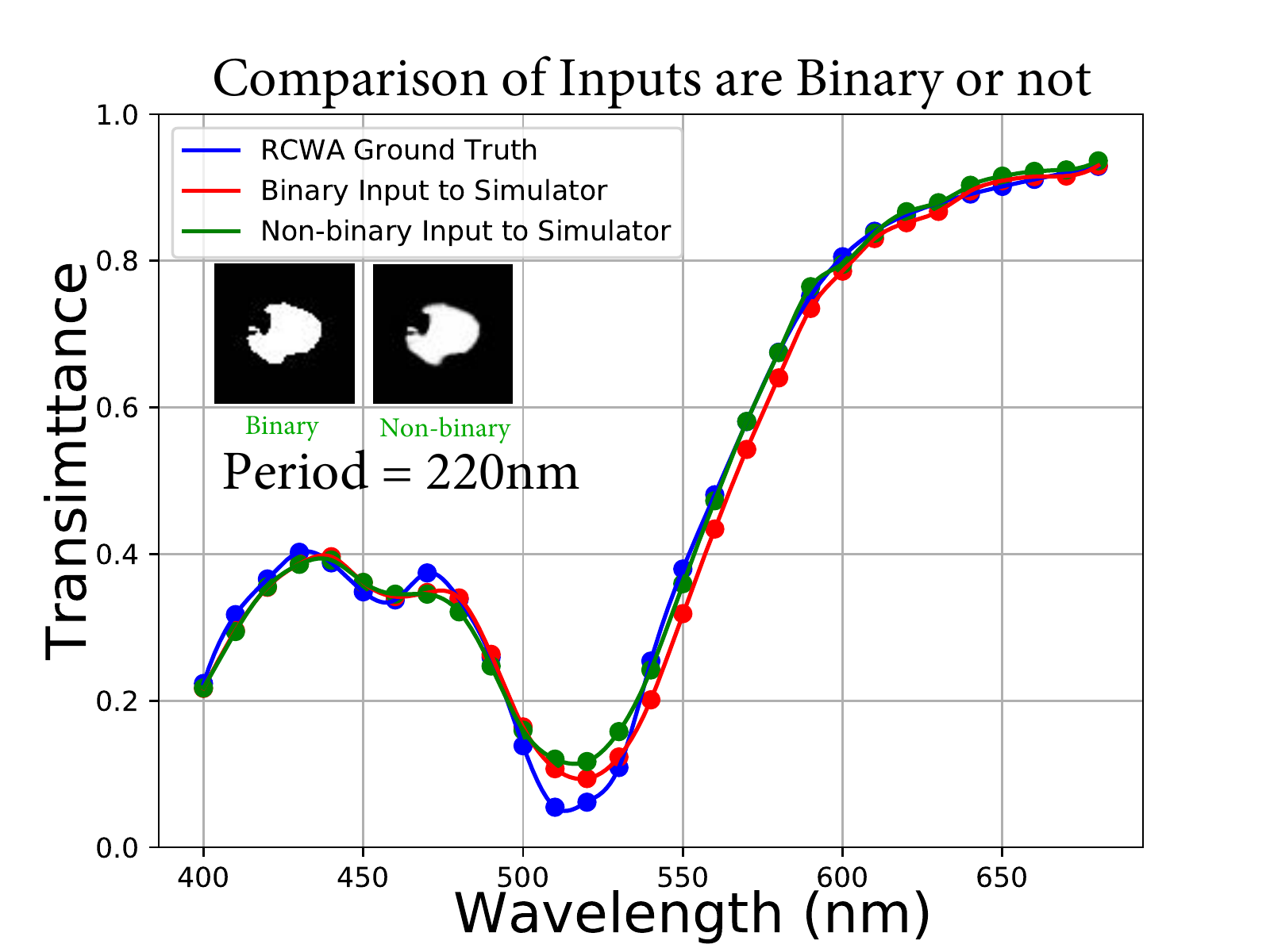}
		\caption{Comparison between the result of RCWA and simulator's output given binary image or not(only TE is shown). The threshold of image binarization is 0.5.}
		\label{dim_vs_clear}
	\end{minipage}
\end{figure*}

Our models perform well on the validation set. However, it is possible that the desired spectrum is different from any one in our data. To illustrate the universality and superiority of our design, we also carry out several comparative experiments. We still use an upside-down Gaussian function (mean=600, variance=40, amplitude=0.9) as the desired spectrum $\bm{T}$ and search for the most similar one by traversing in both our training set and validation set based on the MSE criterion. The detected spectrum with the smallest MSE and the matched device are shown in Figure \ref{c_trad}. To testify the effectiveness of the contrast vector, we train another generator without using contrast vectors while fixing all other training hyperparameters. We feed in the same desired spectrum $\bm{T}$ and the generated device and spectrum are shown in Figure \ref{c_gauss}. Using our well-designed generator mentioned in methods and feeding in the contrast vector $\bm{C}$ converted from $\bm{T}$, the generated device and the spectrum are obtained, plotted in Figure \ref{c_ctrast}. Considering that Polycrystalline Silicon is more lossy for blue light compared with red light, we adjust $\bm{C}$ slightly,  applying minor jitter to every single value while maintaining the overall tendency of the previous vector. For every randomized $\bm{C}$, the smallest value of contrast is 0.01, while the other ones are selected from 0.4, 0.5 or 0.6 with equal probability. Figure \ref{c_random} is the best one of the results obtained by feeding four different revised contrast vectors to the generator. From Figure \ref{c_trad} to Figure \ref{c_random}, we can see that the obtained spectrum is getting closer to $\bm{T}$. It is worth noting that $\bm{T}$ does not exist in our possible design space, according to Figure \ref{pss}. That is why Figure \ref{c_gauss}-\ref{c_random} are not so satisfying as Figure \ref{t_generator}. Nevertheless, our methods still produce an acceptable result. More results from the generator for artificial desired spectra can be found in Figure \ref{more}.

We have found in the above experiments that if the desired spectrum is not a spectrum that real exists, i.e. the input set artificially according to the subjective demand, it is possible that the generated shapes vary greatly from the ones in our training set. For example, in Figure \ref{c_ctrast}, the generated shape has not only one part in the middle but also another isolated one in the upper left corner. It shows that our neural network understands the mapping between devices and spectra during the training process so that it can produce a shape that is totally different from those in the training set.

As mentioned above, we transform the generated shape into a binary array in the process of testing but not implement it in our training process, since the mandatory binarization of an image makes gradients too steep which is not helpful for training. In order to solve this problem, previous work trained a simulator with a noise similar to the generated patterns to circumvent the binarization and smoothing during the predicting process\cite{liu2018generative}. It is less efficient but can ensure that the input image of the simulator is binary and not so complex for manufacture. We purposely experiment to figure out how this post-process affects our results. We feed a non-binary generated image and a binarized version of it to the simulator respectively. Compared with the result given by RCWA, in Figure \ref{dim_vs_clear}, spectra of these two shapes are both very similar to the ground truth. Besides, after training the generator, the mean Manhattan distance of pixels in all generated shapes is approximately 0.01. In other words, the intensity of each pixel ranges from 0 to 0.01 or 0.99 to 1 in an average sense. Thus, binarization can be ignored in the training process.

\begin{figure*}[bt]
	\centering
	\subfloat[]{
		\begin{minipage}{1\linewidth}
			\includegraphics[width=0.1\linewidth]{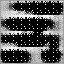}
			\includegraphics[width=0.1\linewidth]{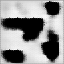}
			\includegraphics[width=0.1\linewidth]{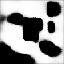}
			\includegraphics[width=0.1\linewidth]{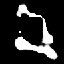}
			\includegraphics[width=0.1\linewidth]{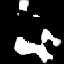}
			\includegraphics[width=0.1\linewidth]{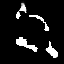}
			\includegraphics[width=0.1\linewidth]{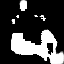}
			\includegraphics[width=0.1\linewidth]{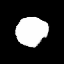}
			\includegraphics[width=0.1\linewidth]{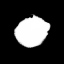}
		\end{minipage}
	}
	
	\subfloat[]{
		\begin{minipage}{1\linewidth}
			\includegraphics[width=0.1\linewidth]{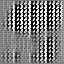}
			\includegraphics[width=0.1\linewidth]{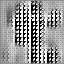}
			\includegraphics[width=0.1\linewidth]{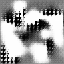}
			\includegraphics[width=0.1\linewidth]{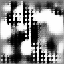}
			\includegraphics[width=0.1\linewidth]{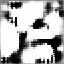}
			\includegraphics[width=0.1\linewidth]{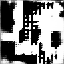}
			\includegraphics[width=0.1\linewidth]{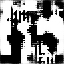}
			\includegraphics[width=0.1\linewidth]{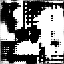}
			\includegraphics[width=0.1\linewidth]{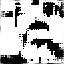}
		\end{minipage}
	}
	\caption{A comparative test on whether SSIM is used as a loss function. For each row, the $n$th image is the predicted shape of generator at the time of $2^{n-1}$ epoch for the same validation input. (a)Predicted shape from generator with SSIM. (b)Predicted shape from generator without SSIM.}
	\label{ssim}
\end{figure*}

Since SSIM is utilized the training process, an unsupervised learning task is changed into a supervised one labelled with shape $\bm{S}$. We also investigate whether SSIM should be applied as a part of the loss function. As shown in Figure \ref{ssim}, SSIM influences both the appearance and the contrast (equals to the degree of binarization here) of the shape generated by the generator, so its existence expedites the network convergence towards a specific direction. Considering that one spectrum can correspond to multiple structures when the generator is in different epochs, it is likely to generate different shapes for the same spectrum. If SSIM is not used, the same loss from distinct shapes will be given to the network in such case, and the network will be confused and difficult to converge.

In the future, we intend to utilize semi-supervised methods to decrease the demands of data. The architecture of neural networks can also be ameliorated and simplified because we observe that the gradients disappear under certain circumstances. Other possible extensions are worth further explorations, such as applying our models in other bands, using more shortcuts and try other descriptive ways other than contrast vectors.

\section{Conclusions}
In summary, for "Forward Simulation" problem, when the required precision is not high, our simulator can substitute the traditional simulation method with a great improvement in efficiency. For the generator in the same given design space, such as the arbitrary shapes of the metasurface and the periods within a specific range, our neural network model can generate a roughly optimal solution for the desired spectrum, even if the desired spectrum does not exist in the design space. Another advantage of our model is that the degree of freedom of the generated device is relatively high, for it is capable of synchronously optimizing the shape and the period of the metasurface. Compared with the traditional traversal method and other artificial intelligence methods, the diversity of our generated devices significantly increases under the premise of ensuring speed and accuracy. In addition to the visible band, our model can also be applied to other bands theoretically. Last but not least, as shown in Figure 3b, our generator can find different structures satisfying almost the same spectrum. Since our generator tends to generate regular patterns, it can be used to improve a relatively complex structure to meet the actual processing requirements better.

The methodology we have developed is readily to be used into migration application for the pursuit of desired complex values of the reflection and
transmission coefficients, which is essential to the metasurface design where the amplitude, the phase and the incident angle of the light waves matter. In the future, our models can be ameliorated and simplified to adapt to problems where multiple one-dimensional parameters work together rather than considering a 
two-dimensional image with single one-dimensional parameter merely. Further subsequent improvement on network performance can also start from incorporation with other deep learning methods, such as using recurrent neural network of natural language processing to extract sequential information from spectrum, imitating the attention mechanism of computer vision to make image generation more explicable. To reduce the burden of generating training data brought by the demand on more parameters, the application of reinforcement learning and unsupervised learning in this field is also vital to mitigate the data's dependency on human prior knowledge. We envision deep learning being widely applied to the optimization of metasurfaces, and even to the entire field of optics, so that scientists and engineers will be greatly relieved from the tedious process of trial and error methods and will focus more on truly creative thinking, just leaving repetitive tasks to machines.

\appendix
\section{Supporting Information}
\subsection{Details about neural network}
\label{appendix}
We generated approximately 6500 pieces of data by RCWA\cite{RETICOLO}. 80 percent of them are training set and 20 percent are validation set. Each piece of data includes a shape, a period and the corresponding spectrum. The spectrum consists of 58 points, two halves of which respresent TE and TM reponse sampled with equally spaced intervals between 400nm and 680nm. The shape is a 64$\times$64 binary image. The period is a integer between 200 and 400nm.

The simulator consists of convolution layers extracting information from images and fully connected layers converting images into vectors. For the simulator, the data is augmented before being fed. Adaptive moment estimation (Adam)\cite{kingma2014adam} is used to update the gradient. Learning rate becomes smaller from 0.02 as the epoch becomes larger. We use a Telsa K80 GPU to train simulator with 500 epochs for approximately half an hour.

The generator consists of the deconvolution layers to generate images from sequences and the fully connected layers to obtain features from images. The shape is generated first after deconvolution layers, then the period is produced after fully connected layers. Besides, we use a shortcut to make the training process more stable and fast. The input noise is sampled from a uniform distribution between 0 and 1. Adam is used to update the gradient, and the learning rate becomes smaller from 0.02 as the epoch becomes larger. We use a Telsa K80 to train the generator with 1000 epochs for approximately an hour.

\begin{figure*}[bt]
	\centering
	\subfloat[]{
		\begin{minipage}{0.48\linewidth}
			\includegraphics[width=1\linewidth]{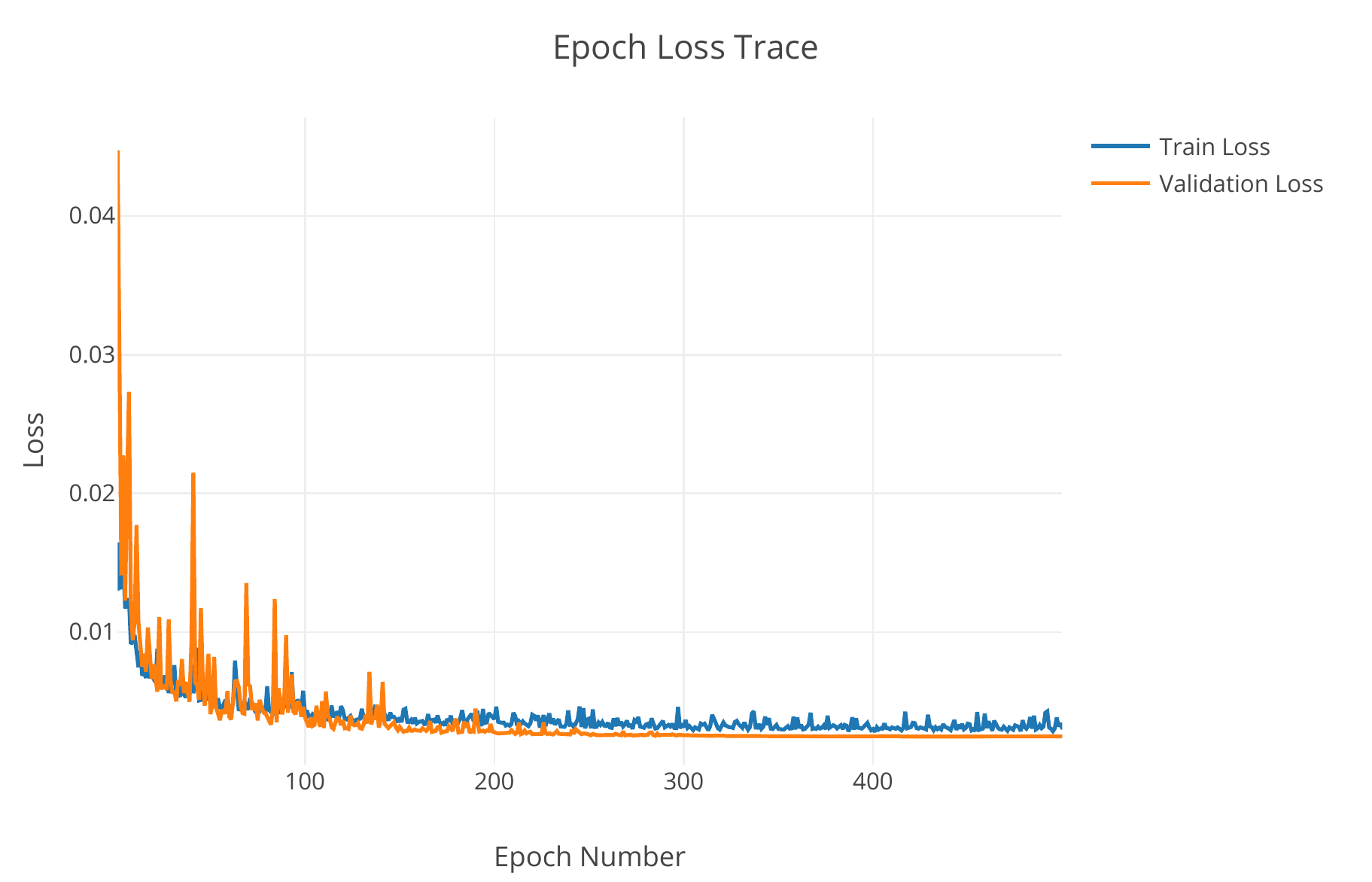}
		\end{minipage}
	}
	\subfloat[]{
		\begin{minipage}{0.48\linewidth}
			\includegraphics[width=1\linewidth]{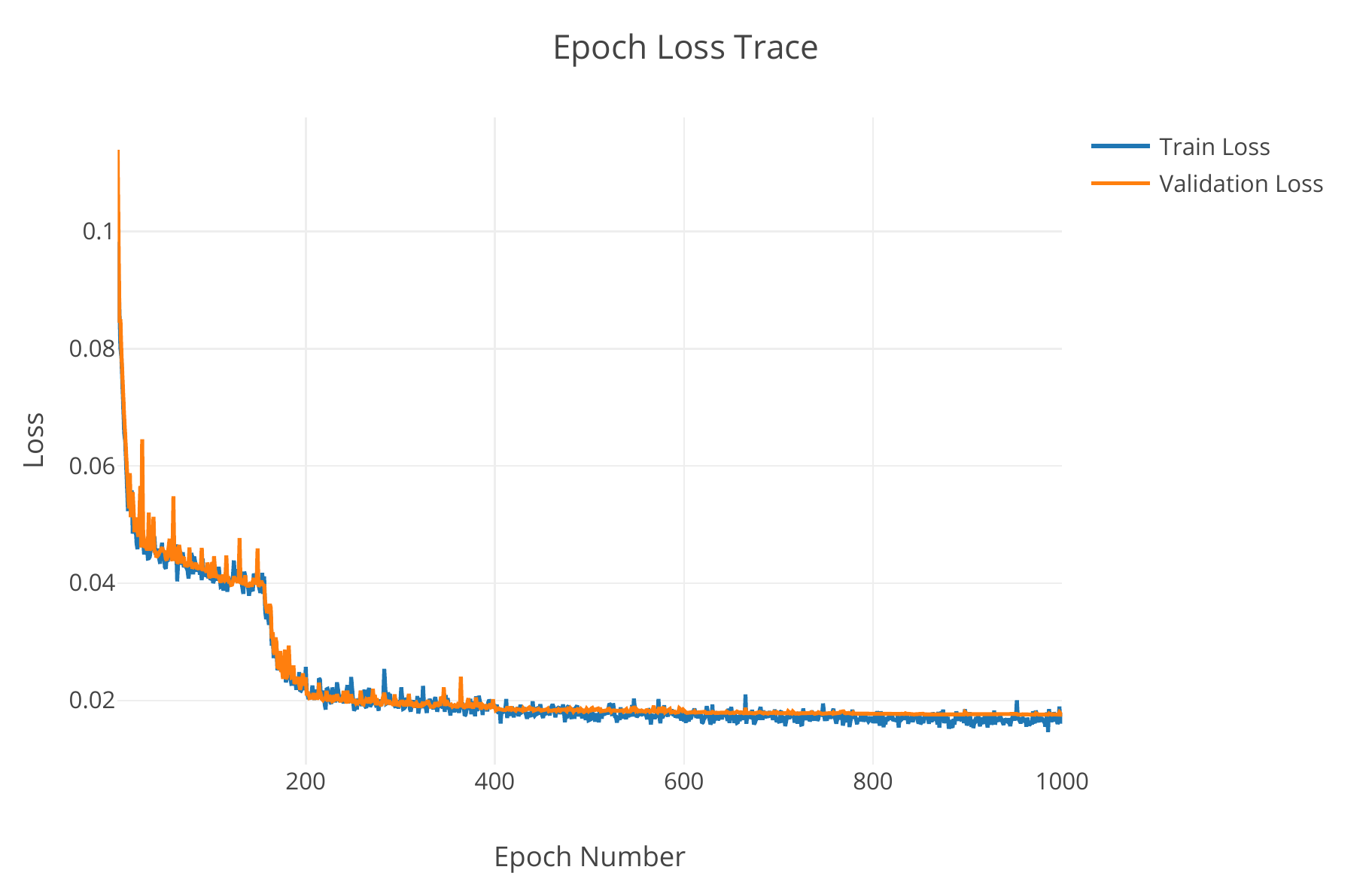}
		\end{minipage}
	}
	\caption{(a) Loss curves of the simulator. (b) Loss curves of the generator.}
	\label{loss}
\end{figure*}

The detailed hyperparameters we used are listed in Table \ref{hyper}, and loss curves of simulator and generator are shown in Figure \ref{loss}.

\begin{table*}[bt]
	\caption{Hyperparameters used in the training of the simulator and the generator.}
	\centering
	\renewcommand\arraystretch{2}
	\begin{tabular}{|c|c|c|}
		\hline
		& \textbf{Simulator}              & \textbf{Generator}              \\ \hline
		\textbf{Epoch}            & 500                             & 1000                            \\ \hline
		\textbf{Batch size}       & 1024                            & 256                             \\ \hline
		\textbf{Optimizer}        & Adam ($\beta_1$=0.5, $\beta_2$=0.999)& Adam ($\beta_1$=0.5, $\beta_2$=0.999)\\ \hline
		\textbf{Learning rate}    & Initial=0.02, Step=100, $\gamma$=0.5 & Initial=0.02, Step=200, $\gamma$=0.5 \\ \hline
		\textbf{Initial strategy} & Mean=0, Std=0.02                 & Mean=0, Std=0.02                 \\ \hline
		\textbf{Loss parameter}   & None                             & $\alpha$=0.05, $\beta$=0         \\ \hline
	\end{tabular}
	\label{hyper}
\end{table*}

\begin{table*}[bt]
	\caption{The mapping relationship between color and function.}
	\centering
	\renewcommand\arraystretch{2}
	\begin{tabular}{|c|c|c|c|c|}
		\hline
		\textbf{Color}    & Yellow        & Cyan                  & Lavender        &         \\ \hline
		\textbf{Function} & Convolution   & Transpose Convolution & Full Connection &         \\ \hline
		\textbf{Color}    & Blue          & Orange                & Purple          & Red     \\ \hline
		\textbf{Function} & Normal Vector & Leaky ReLu (p=0.2)    & Tanh            & Sigmoid \\ \hline
	\end{tabular}
	\label{mapping}
\end{table*}

\subsection{Algorithm for generating random structures}
\url{https://stackoverflow.com/questions/8997099/algorithm-to-generate-random-2d-polygon}

\subsection{Contrast vectors}
If we use the spectra with 58 points instead of the contrast vectors, the loss curve of the generator suggests that the spectrum loss is smaller, approximately 4\% error rate. Nevertheless, if we feed an artificial spectrum to it, the output results deteriorate a lot. As we discussed in the paper, the suboptimal solution cannot be found because the network aims at minimizing MSE ultimately. In practice, the network pays more attention to the contrasts instead of inconsequential details of a spectrum, thus leading to a better result. Besides, we are indifferent to the detailed features of the desired spectrum. So we need a more intuitive method to describe it. It makes senses that the use of contrast vectors is essentially a trade-off for universality at the expense of network prediction accuracy.

\bibliography{hfid}

\end{document}